\documentclass[aps,prl,showpacs,twocolumn,superscriptaddress]{revtex4-1}
\usepackage{amsmath,amssymb,amsfonts,bm}
\usepackage{graphicx}
\usepackage{epstopdf}
\usepackage{dcolumn}
\usepackage{mathrsfs}
\usepackage{bbold}
\usepackage{dsfont}
\usepackage{dcolumn}

\usepackage[colorlinks=true,linkcolor=blue,citecolor=blue, urlcolor=blue,bookmarks=false]{hyperref}

\begin{document}

\title{Higher-Order Topological Insulators in Quasicrystals}
\author{Rui Chen}
\affiliation{Department of Physics, Hubei University, Wuhan 430062, China}
\author{Chui-Zhen Chen}
\affiliation{Institute for Advanced Study and School of Physical Science and Technology, Soochow University, Suzhou 215006, China}
\author{Jin-Hua Gao}
\thanks{jinhua@hust.edu.cn}
\affiliation{School of Physics and Wuhan National High Magnetic Field Center, Huazhong University of Science and Technology, Wuhan 430074, China}
\author{Bin Zhou}\thanks{binzhou@hubu.edu.cn}
\author{Dong-Hui Xu} \thanks{donghuixu@hubu.edu.cn}
\affiliation{Department of Physics, Hubei University, Wuhan 430062, China}

\begin{abstract}
Current understanding of higher-order topological insulators (HOTIs) is based primarily on crystalline materials. Here, we propose that HOTIs can be realized in quasicrystals. Specifically, we show that two distinct types of second-order topological insulators (SOTIs) can be constructed on the quasicrystalline lattices (QLs) with different tiling patterns. One is derived by using a Wilson mass term to gap out the edge states of the quantum spin Hall insulator on QLs. The other is the quasicrystalline quadrupole insulator~(QI) with a quantized quadrupole moment. We reveal some unusual features of  the corner states~(CSs) in the quasicrystalline SOTIs. We also show that the quasicrystalline QI can be simulated by a designed electrical circuit, where the CSs can be identified by measuring the impedance resonance peak. Our findings not only extend the concept of HOTIs into quasicrystals but also provide a feasible way to detect the topological property of quasicrystals in experiments.
\end{abstract}

\maketitle
\emph{Introduction.}---Since the discovery of topological insulators~(TIs)~\cite{Hasan2010RMP,Qi2011RMP}, tremendous effort has been devoted into the search for exotic topological phases~(TPs) of matter. Recently, higher-order topological insulators~(HOTIs)~\cite{Benalcazar2017Science} were proposed as an extension of TIs, which have been widely investigated in condensed matter as well as phononic, microwave, photonic and electrical circuit~(EC) systems~\cite{Langbehn2017PRL,Benalcazar2017PRB,Song2017PRL,Schindler2018SciAdv, Ezawa2018PRL,Ezawa2018PRL1,Schindler2018NatPhys,Liu2019PRL,Serra_Garcia2018Nature,Peterson2018Nature,Noh2018NatPho,Imhof2018NatPhys,Ni2018NatMat,Xue2018NatMat,vanMiert2018PRB,Ezawa2018PRB,Ezawa2018PRB1,Xie2018PRB,Kunst2018PRB,FrancaPRB2018,Matsugatani2018PRB,Khalaf2018PRB,Geier2018PRB,Zhang2019NatPhys,Liu2019PRL1,Trifunovic2019PRX,Xie2019PRL,Chen2019PRL,Serra_Garcia2019PRB,Lee2019PRL}. Unlike conventional TIs in $d$ dimensions which have gapless states on the $d-1$-dimensional boundary, $n$th-order ($1< n \leq d$) TIs in $d$ dimensions have $(d-n)$-dimensional gapless boundary states~\cite{Benalcazar2017Science,Langbehn2017PRL,Benalcazar2017PRB,Song2017PRL,Schindler2018SciAdv}. For instance, two-dimensional~(2D) second-order topological insulators~(SOTIs) display robust zero-energy modes~(ZEMs) localized at their $0$-dimensional corners, dubbed corner states~(CSs).

 The study of TPs has been lately extended to aperiodic quasicrystalline systems~\cite{Huang2018PRL,Huang2018PRB,Bandres2016PRX,Tran2015PRB,Fulga2016PRL,Kraus2012PRL}, which lack translational symmetry and show forbidden symmetries in crystals such as the 5-fold and 8-fold rotation symmetries. The proposals for realizing conventional TIs~\cite{Huang2018PRL,Huang2018PRB,Tran2015PRB,Bandres2016PRX} have been proposed in quasicrystalline systems. It is of interest to ask if it is possible to realize HOTIs in quasicrystals.

In this Letter, we propose SOTIs on 2D quasicrystalline lattices~(QLs), which extend the concept of HOTIs to quasicrystalline systems. We demonstrate two general schemes to construct SOTIs in 2D quasicrystals. First, an SOTI can be realized by adding a proper mass term into the quantum spin Hall~(QSH) insulator on a QL, where the edge states are gapped, and then topological CSs emerge. Second, the other type of SOTIs in quasicrystals originates from a quantized bulk quadrupole moment~\cite{Benalcazar2017Science,Benalcazar2017PRB}, which is a quasicrystalline quadrupole insulator~(QI). With the two mechanisms, the hallmark zero-energy CSs are found on the designed QLs. Our results reveal several distinguishing features of the quasicrystalline SOTIs: (1) the CSs strongly rely on the tiling pattern and boundary geometry of quasicrystals; (2) intriguing extended ZEMs exist in quasicrystals, quite unlike the normal CSs, they distribute along some segments of the edge; (3) 8-fold symmetric CSs are found in the Ammann-Beenker~(AB) tiling quasicrystal, which are protected by a rotation symmetry $C_8$ forbidden in crystals, indicating that the quasicrystalline HOTI is actually distinct from that in crystalline systems. We also design an EC to simulate the quasicrystalline QI and show that the CSs can be measured as an impedance resonance peak at the EC corners~\cite{Imhof2018NatPhys}. It may be the first feasible way to experimentally detect quasicrystalline HOTIs.

\emph{Mass term induced SOTIs in quasicrystals.}---It was known that QSH states can be realized on 2D QLs~\cite{Huang2018PRL}. The first scheme to make an SOTI in a 2D quasicrystal is to use an additional mass term to gap the topological edge states. Once this mass term results in a domain wall structure at two adjacent edges, a CS appears at the intersection of the two edges, \textit{i}.\textit{e}., an SOTI on the QL is achieved.

\begin{figure}[t]
	\includegraphics[width=8cm]{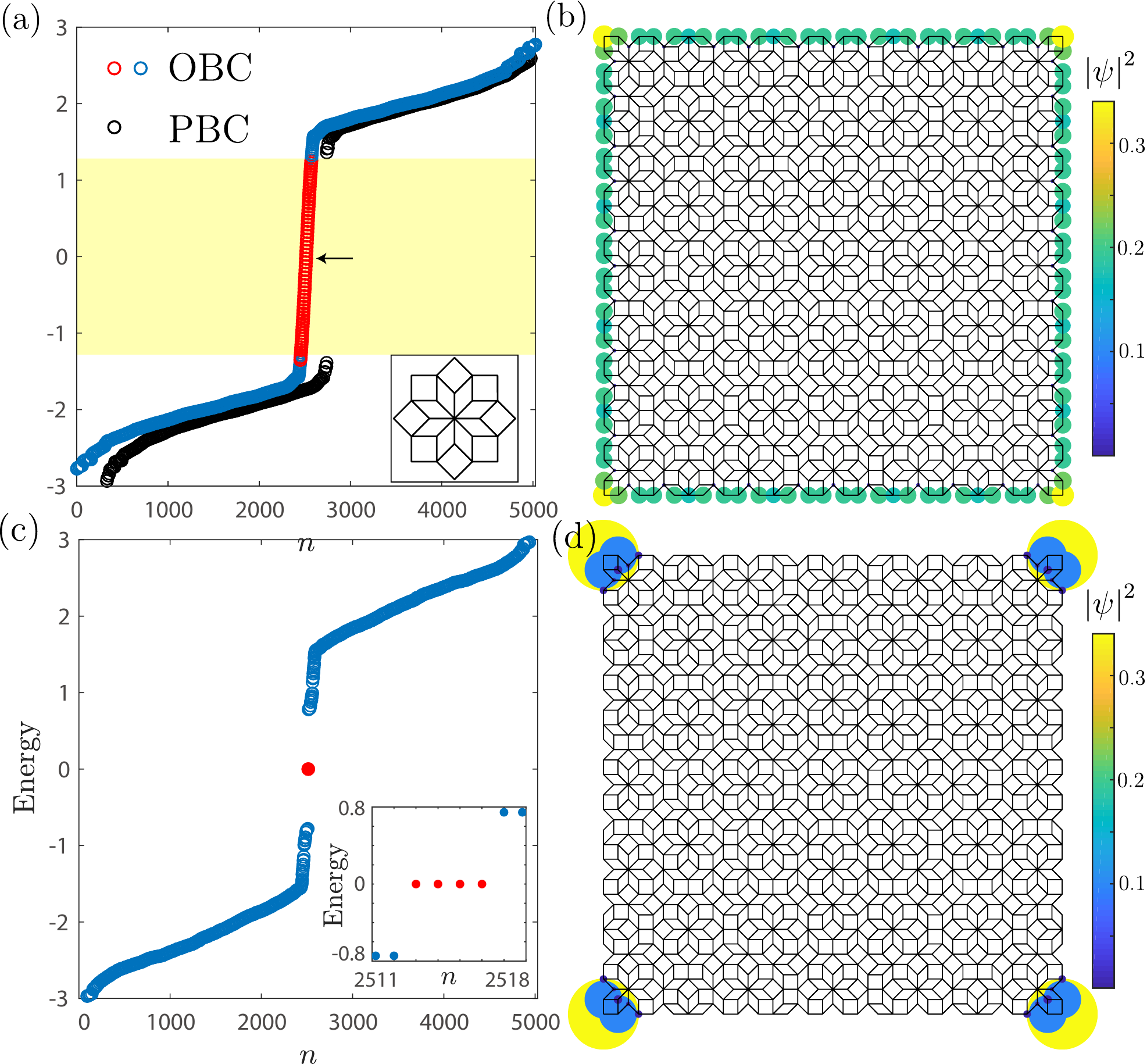}
	\caption{(a) Energy spectra of $H_\text{QSH}$ in the AB tiling quasicrystals with PBC and OPC versus the eigenvalue index $n$. The inset denotes a small pattern that repeats itself throughout the QL. (b) The wavefunction probability of the in-gap state marked by a black arrow in (a). (c) Spectrum of the SOTI on the AB tiling QL with the OBC versus $n$. The inset shows the zoomed-in section of four ZEMs marked as the red dots. (d) The probability of the ZEMs in (c). The number of lattices is 1257.}
	\label{fig1}
\end{figure}

We first design a QSH insulator on a QL constructed according to the AB tiling with 8-fold rotation symmetry, where the plane is tiled using squares and rhombi as shown in Fig.~\ref{fig1}. In our model, each lattice site has two orbitals and the Hamiltonian is
\begin{eqnarray}
\label{model1}
H_\text{QSH}&=&-\sum_{j\neq k}\frac{f(r_{jk})}{2} c_{j}^{\dag }\big[it_{1}\left( s _{3}\tau _{1}\cos\phi_{jk}+s _{0}\tau _{2}\sin\phi_{jk}\right)\nonumber\\
&+&t_{2}s _{0}\tau_{3}\big] c_{k}+\sum_{j} \left(M+2t_{2}\right)c_{j}^{\dag }s _{0}\tau _{3} c_{j},
\end{eqnarray}
where $c^\dagger_{j}=(c^\dagger_{j\alpha\uparrow},c^\dagger_{j\alpha\downarrow},c^\dagger_{j\beta\uparrow},c^\dagger_{j\beta\downarrow})$ are electron creation operators at site $j$. $\alpha$ and $\beta$ represent two orbitals at one site and spin degrees of freedom are considered. $M$ denotes the Dirac mass, $t_1$ and $t_2$ are hopping amplitudes. $s_0$ is the $2\times2$ identity matrix, $s_{1,2,3}$ and $\tau_{1,2,3}$ are the Pauli matrices acting on the spin and orbital spaces, respectively. $\phi_{jk}$ is the polar angle of bond between site $j$ and $k$ with respect to the horizontal direction~\cite{Fulga2016PRL,Agarwala2019arXiv}. $f\left( r_{jk}\right) = e^{1-r_{jk}/\xi}$ is the spatial decay factor of hopping amplitudes with the decay length $\xi$ and $r_{jk}=\left|\mathbf{r}_j-\mathbf{r}_k\right|$.
$H_\text{QSH}$ preserves time-reversal~$T$, particle-hole~$C$ and chiral~$S=TC$ symmetries, where $T=is_2\tau_{0}K$, and $C=s_3\tau_1K$ with complex conjugate $K$, therefore it belongs to the class DIII~\cite{Schnyder2008PRB,Kitaev2009AIP,Chiu2016RMP}. In some sense, this QSH model can be understood as a derivative of the Bernevig-Hughes-Zhang model~\cite{Bernevig2006Science}, which describes the QSH states in HgTe quantum wells.
In the following calculations, we fix the side length of rhombi or squares $c=1$ and include hoppings up to the third nearest neighbors.

With the designed QL above, we get a QSH state as shown in Figs.~\ref{fig1}(a) and \ref{fig1}(b). We set $\xi=1$, $M=-1$ and $t_1=t_2=1$, and diagonalize $H_\text{QSH}$ under a periodic boundary condition~(PBC) and an open boundary condition~(OBC). The energy spectrum in Fig.~\ref{fig1}(a) exhibits an energy gap for the PBC, while for the OBC, gapless in-gap states occupy the bulk gap. Figure~\ref{fig1}(b) displays the wavefunction probability of a typical in-gap state, which suggests a localized edge state. Due to time-reversal symmetry~(TRS), each in-gap energyvalue is doubly degenerate, corresponding to two counter-propagating edge states, which characterize the QSH state. Note that the QLs, which can host QSH states, are not unique. Recently, the QSH insulator on the Penrose tiling QL was also proposed~\cite{Huang2018PRL,Huang2018PRB}.
\begin{figure*}[t]
\includegraphics[width=14cm]{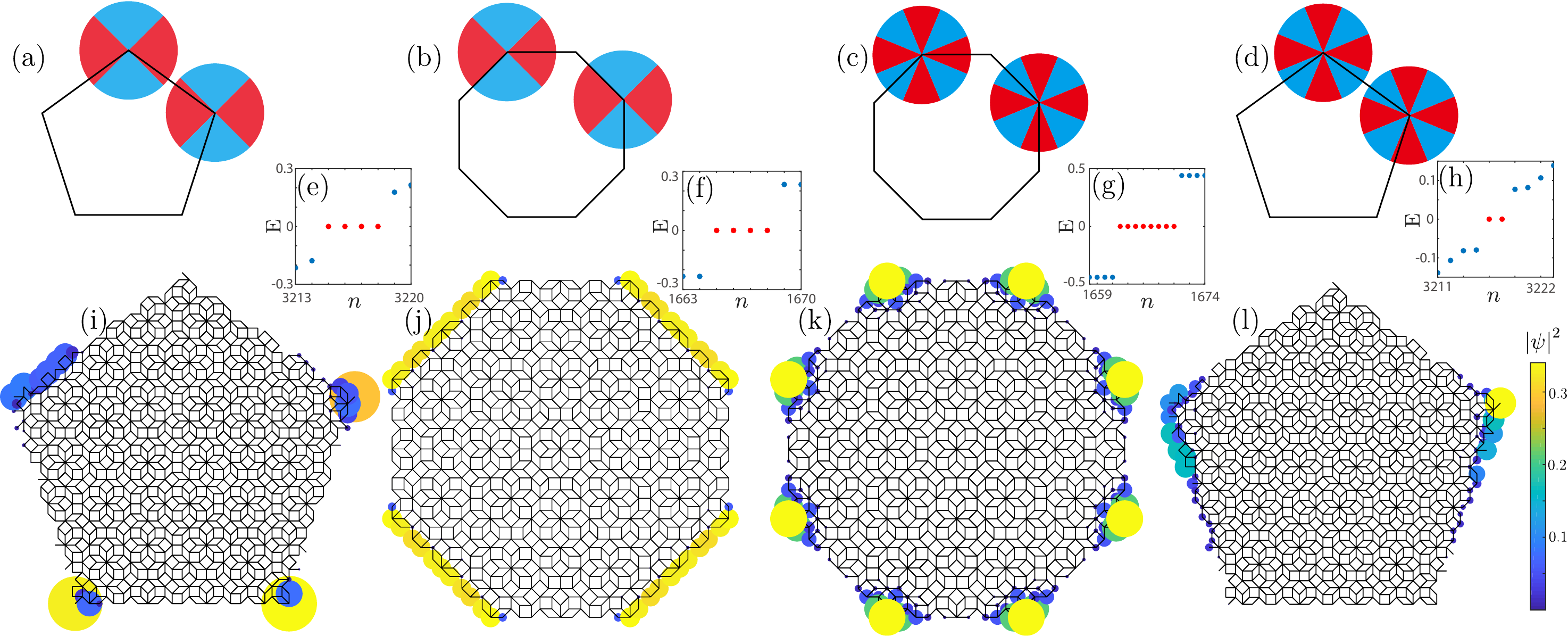}
\caption{Energy spectra and wavefunction probability of ZEMs in the AB tiling quasicrystals with pentagon and octagon boundaries. In (a-d), the red and blue regions in the color circles denote the two regions of the edge orientation with opposite sign of the effective mass. (e-h) Energy spectra near zero versus $n$. The red dots mark the ZEMs. (i-l) The probability of the ZEMs shown in (e-h). The number of lattices in (i) and (l) is 1608, and in (j) and (k) is 833. $\eta=2$ in (a), (b), (e), (f), (i) and (j), and $\eta=4$ in (c), (d), (g), (h), (k) and (l).}
\label{fig2}
\end{figure*}

To transform the QSH insulator on the QL to an SOTI, we induce a TRS breaking Wilson mass term, which reads
\begin{equation}\label{mass}
H_\text{m}(\eta)=g\sum_{j\neq k}\frac{f(r_{jk})}{2}\cos\left( \eta\phi_{jk} \right) c_{j}^{\dag }s _{1}\tau _{1}  c_{k},
\end{equation}
where $g$ and $\eta$ describe the magnitude and the varying period of the Wilson mass, respectively. Thus, the total Hamiltonian of the SOTI is $H_\text{I}=H_\text{QSH}+H_\text{m}$. In the following, we fix $g=1$ unless otherwise specified.

When turning on $H_\text{m}$, the edge states are gapped out and then the hallmark CSs of HOTIs emerge. Figure~\ref{fig1}(c) gives the energy spectrum of the QL with a square boundary when $\eta=2$, we can see that four ZEMs emerge inside the edge gap and are symmetrically localized at the four corners of the QL~[Fig.~\ref{fig1}(d)]. 
Essentially, the location of CSs can be explained by the Jackiw-Rebbi mechanism~\cite{Jackiw1976PRD} that a topological ZEM appears when a mass domain wall forms. Note that the Wilson mass term in Eq.~\eqref{mass} depends on the polar angle of the bond $\phi_{jk}$. However, due to the lack of translation symmetry in quasicrystals, we can not give an analytic expression of the effective Wilson mass for edge states. A rough approximation is to consider an edge of the sample boundary as a long ``bond", so that the sign of the effective Wilson mass for the edge state depends on the orientation of the edge (or the polar angle of the edge $\theta_\text{edge}$). Surprisingly, this approximation works quite well compared with the numerical results. It can even be used as an intuitive rule to determine the appearance of CSs in any quasicrystal polygons.



In Fig.~\ref{fig2}, we consider quasicrystal pentagon and octagon, which are constructed in the same way as Fig.~\ref{fig1} but of different boundary shapes. We give the results of the quasicrystal pentagon with $\eta=2$ in Figs.~\ref{fig2}(a), \ref{fig2}(e) and \ref{fig2}(i). At all the corners of the pentagon, in-gap ZEMs are found except the top one, and the distribution of the ZEMs become asymmetric [see Figs.~\ref{fig2}(e) and \ref{fig2}(i)]. Generally speaking, the factor $\cos{\eta \theta_\text{edge}}$ in the Wilson mass will distinguish two different regions of the edge orientation $\theta_\text{edge}$, which have opposite sign of the Wilson mass. For example, the two regions for $\eta=2$ are illustrated in Figs.~\ref{fig2}(a) and \ref{fig2}(b), where the red region is determined by $\theta_\text{edge} \in (-\frac{\pi}{4},\frac{\pi}{4})\cup(\frac{3\pi}{4},\frac{5\pi}{4})$, and the blue region is $\theta_\text{edge} \in (-\frac{\pi}{4},-\frac{3\pi}{4})\cup(\frac{\pi}{4},\frac{3\pi}{4})$. For the top corner of the pentagon, the two edges are both in red regions [Fig.~\ref{fig2}(a)], which means that their masses are of the same sign. This is the reason why there is no ZEM at this corner. Meanwhile, for the other four corners, the two edges of each corner lie in two different regions, so that a ZEM appears at each corner. 
The case of the quasicrystal octagon [Figs.~\ref{fig2}(b), \ref{fig2}(f), and \ref{fig2}(j)] is special. Note that the Wilson mass at four edges of the octagon is zero because the polar angles of the four edges are $\theta_\text{edge}=\pm \frac{\pi}{4}, \pm \frac{3\pi}{4}$. At the same time, the remaining four edges of the octagon have a finite Wilson mass. In this situation, we find four in-gap ZEMs distributing along the four edges with zero Wilson mass [Figs.~\ref{fig2}(f) and \ref{fig2}(j)], quite unlike the normal CSs. To the best of our knowledge, these extended ZEMs in the quasicrystal haven't been reported in other HOTIs.

To further confirm the topological origin of CSs, we first examine robustness of the CSs by applying various symmetry-breaking perturbations~\cite{Supplement}. It's found that the CSs in the quasicrystal square~(Fig.~1) are protected by the combined symmetry $C_{4}m_z$ and $C$ symmetry, where  $C_4=e^{-i\frac{\pi }{4}s_{3}\tau _{3}}\mathcal{R}_{4}$ is a 4-fold rotation symmetry and $m_z=s_{3}\tau _{0}$ is the mirror symmetry about the $x-y$ plane. Moreover, the CSs in the AB tiling quasicrystalline HOTI can be characterized by a $\mathbb{Z}_2$ topological invariant, which is determined by the product of Pfaffians at high-symmetry momenta in a 4D momentum hyperspace \cite{DanielVarjas2019arXiv}. $\mathbb{Z}_2$ phase diagram of the SOTI model is shown in Ref.~\cite{Supplement}. Note that the symmetry of a quasicrystal sample also depends on its boundary shape. Thus, the CSs of the quasicrystal pentagon in Fig.~2 (i) aren't protected by the above spatial symmetry. In this case, we actually realize an ``extrinsic'' HOTI \cite{Geier2018PRB}, which hosts termination-dependent corner states instead of spatial-symmetry-protected ones.


We can also change the angular dependence of the Wilson mass by choosing a different $\eta$. With $\eta=4$, We calculate the quasicrystal octagon [Figs.~\ref{fig2}(c), \ref{fig2}(g), and \ref{fig2}(k)] and pentagon [Figs.~\ref{fig2}(d), \ref{fig2}(h), and \ref{fig2}(l)]. Now the red and blue regions for the edge orientation to determine the sign of Wilson mass have been changed [Figs.~\ref{fig2}(c) and \ref{fig2}(d)]. The red region becomes $\theta_\text{edge} \in \cup(-\frac{\pi}{8}+\frac{n\pi}{4},\frac{\pi}{8}+\frac{n\pi}{4})$, where $n=0,1,2,3$. In a similar way, we can predict that the quasicrystal octagon has eight CSs and the quasicrystal pentagon has two corners. It's in good agreement with the numerical results as shown in Figs.~\ref{fig2} (g), \ref{fig2}(k), \ref{fig2}(h), and \ref{fig2}(l).
For $\eta=4$, the CSs of quasicrystal octagon are protected by the combined symmetry $C_8m_z$ and $C$ symmetry. Similarly, we can also define a $\mathbb{Z}_2$ topological invariant, and the $\mathbb{Z}_2$ phase diagram for $\eta=4$ is given in Ref.~\cite{Supplement}. We emphasize that the 8-fold symmetric CSs here are forbidden in crystals, as crystals don't have a $C_8$ symmetry. This implies that the quasicrystalline HOTI is beyond the framework of crystalline HOTIs.


The results above indicate that, with an angular dependent Wilson mass term, SOTIs can be achieved on the constructed QLs, and CSs can be manipulated by choosing proper boundaries (See Ref.~\cite{Supplement} for other geometries). It should be emphasized that we can also use other kinds of tilings, e.g., the Penrose tiling~\cite{Supplement}, to construct a quasicrystalline SOTI.

\begin{figure}[t]
\includegraphics[width=8cm]{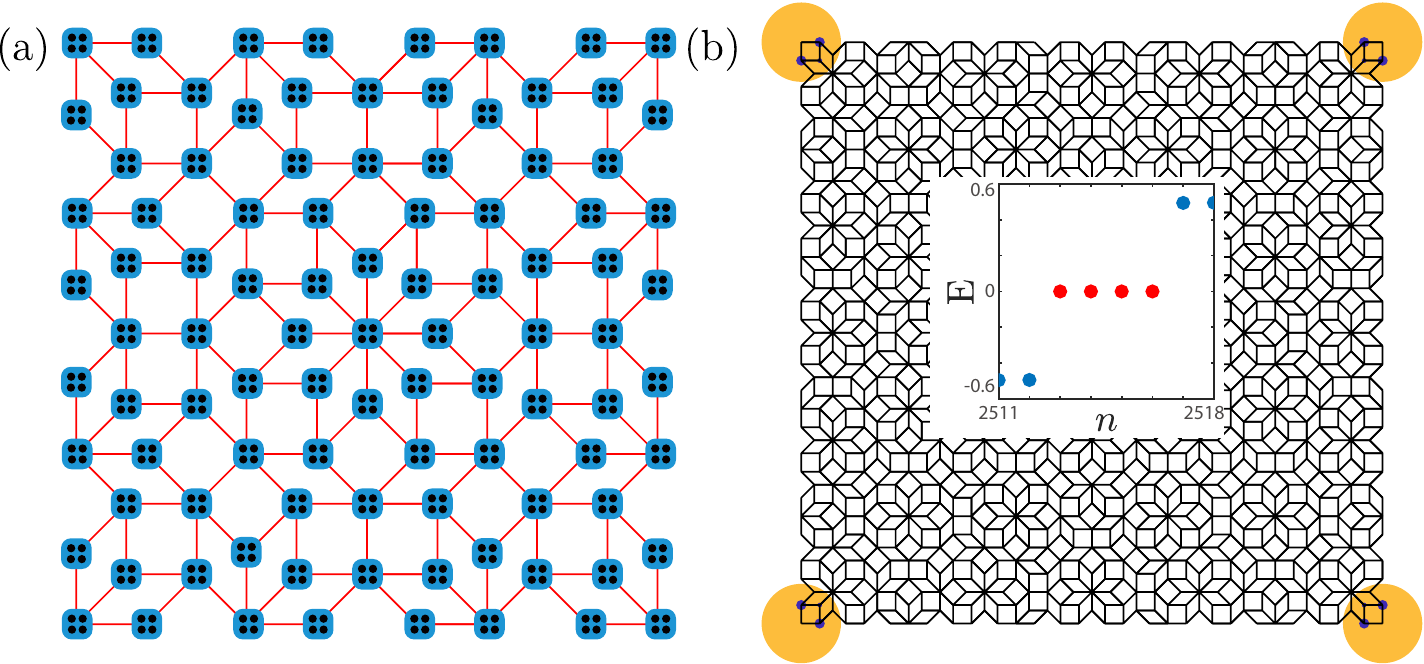}
\caption{(a) Schematic drawing of the AB tiling QL. Shaded blue squares mark the cells. (b) The probability of the ZEMs. The inset shows the spectrum near zero. Red dots mark the four ZEMs. The number of cells is 1257.}
\label{fig3}
\end{figure}

\emph{QIs in quasicrystals.}---The other feasible scheme to construct an SOTI on a QL is to consider the QI with a quantized quadrupole moment. Referring to the 2D square lattice model of QI in Refs.~\cite{Benalcazar2017Science, Benalcazar2017PRB}, we design a 2D QL as shown in Fig.~\ref{fig3}(a). Here, four sites form a cell, and we use the cells to construct the AB tiling QL. The Hamiltonian is
\begin{equation} \label{model2}
H_\text{QI}\!=\!\gamma \sum_{j}c^\dagger_j\left( \Gamma _{2}\!+\!\Gamma_{4}\right) c_j
\!+\!\frac{\lambda }{2}\sum_{j\neq k} f\left(r_{jk}\right) c_{j}^{\dag }T(\phi_{jk})c_{k},
\end{equation}
with
$T(\phi_{jk})= \left\vert \cos \phi_{jk} \right\vert \Gamma _{4}-i\cos \phi_{jk} \Gamma_{3}+\left\vert \sin \phi_{jk} \right\vert \Gamma _{2}-i\sin\phi_{jk} \Gamma _{1}.$
Here, $c^\dagger_j=(c^\dagger_{j1},c^\dagger_{j2},c^\dagger_{j3},c^\dagger_{j4})$ is the electron creation operator in cell $j$. The first and second terms in Eq.~\eqref{model2} are the intra-cell and inter-cell hoppings with amplitudes $\gamma$ and $\lambda$.
 $\Gamma _{4}=\tau_{1}\tau _{0}$ and $\Gamma _{\nu}=-\tau_{2}\tau_{\nu}$ with $\nu=1,2,3$.  $\tau_{1,2,3}$ are the Pauli matrices representing the sites in one cell, and $\tau_{0}$ is the identity matrix.

We consider the AB tiling QL with a square boundary and include the hoppings to the third neighbors. By numerically diagonalizing $H_\text{QI}$ with $\gamma=0.1\lambda$, we find four zero-energy CSs on the designed QL in Fig.~\ref{fig3}(b). This system has two mirror symmetries $m_x$ and $m_y$ as well as a rotation symmetry $C_4$, resulting in a quantized quadrupole moment $Q_{xy}=0,e/2$. Thus, $Q_{xy}$ is a natural topological invariant and can be calculated in real space~\cite{Wheeler2018arXiv,Kang2018arXiv}. Numerical calculations show that corner states appear when $Q_{xy}=e/2$~\cite{Supplement}. These results clearly indicate that Eq.~\eqref{model2} on the designed QL can produce a QI. The results of the Penrose tiling QL are presented in Ref.~\cite{Supplement}.

\emph{Electrical-circuit realization.}---
 Here, we show that the QL hosting the QI we proposed can be mapped onto an EC lattice, and the zero-energy CSs will result in an impedance resonance peak at the EC corners.

\begin{figure}[t]
\includegraphics[width=8cm]{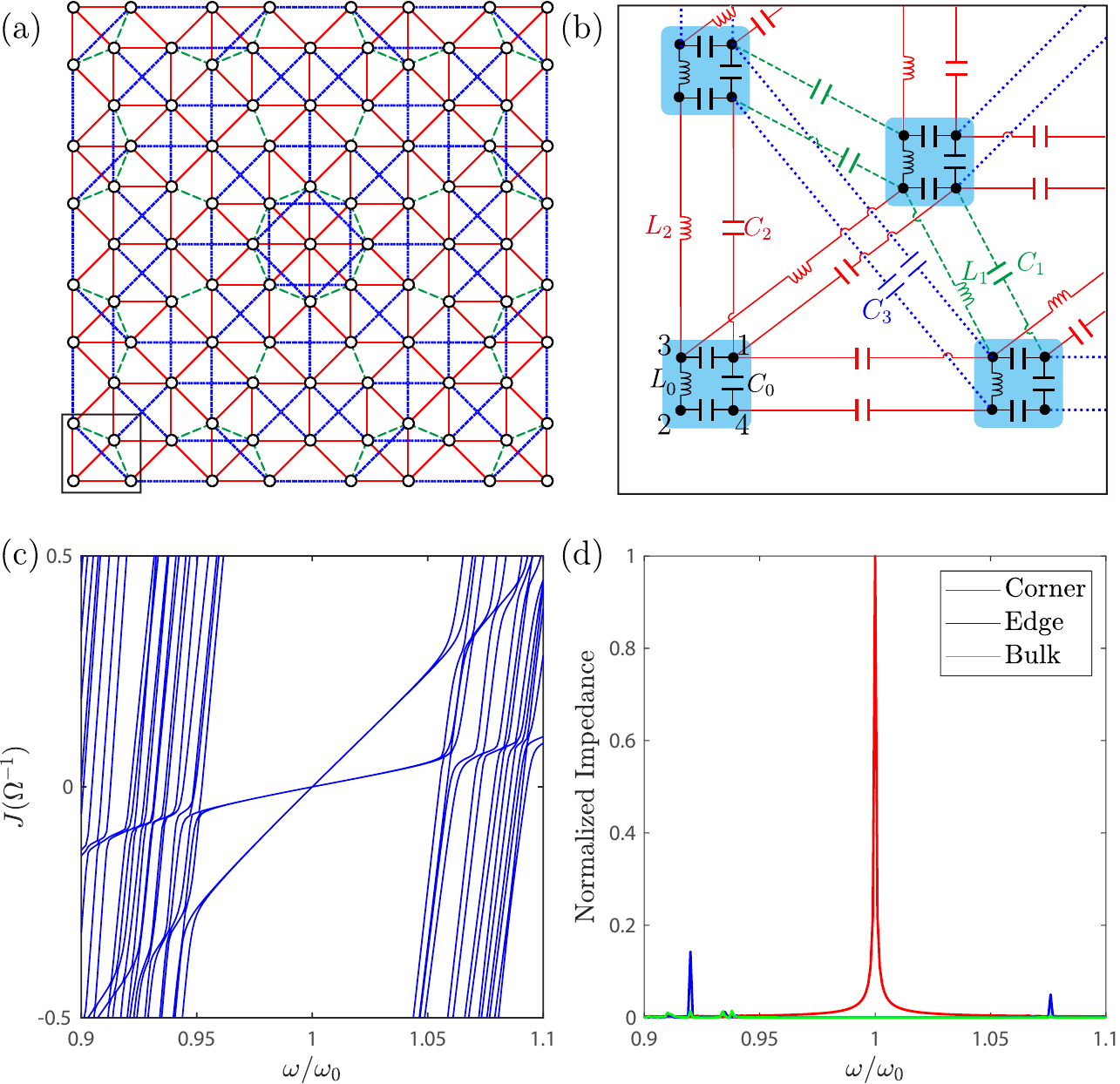}
\caption{(a) Quasicrystalline EC lattice. White dots represent the circuit cells, and the green dashed, red solid, and blue dotted lines correspond to different inter-cell connections. (b) Layout of the capacitors and inductors in the black square in (a). (c) Spectrum of Laplacian $J$ versus $\omega/\omega_{0}$. (d) $Z\left( \omega \right) $ between two nearest-neighbor nodes at the corner, the edge and in the bulk. Here, $C_0=1\text{nF}$ and $L_0=1\mu\text{H}$.}
\label{fig4}
\end{figure}

In a recent work~\cite{Imhof2018NatPhys}, 2D crystalline QI has been simulated by an EC, and the CSs have been observed in experiments. The case of the quasicrystalline QI is rather similar. The designed EC is shown in Figs.~\ref{fig4}(a) and \ref{fig4}(b). Each node in the EC corresponds to a QL site. Similar as the lattice configuration in Fig.~\ref{fig3}, we use four nodes to form a cell [Fig.~\ref{fig4}(b)], and the cells are connected to produce an AB tiling EC [Fig.~\ref{fig4}(a)]. The inter-cell and intra-cell connections between the nodes with capacitors or inductors are given in Fig.~\ref{fig4}(b), and more details are presented in~\cite{Supplement}. Here, we make a simplification about the inter-cell hopping on the QL and set $T\left( \phi_{jk}\right)= \left( 1-n \right)\left[\Gamma_4-i\left(-1\right)^m\Gamma_3\right] + n\left[ \Gamma_2-i\left(-1\right)^m\Gamma_1\right]$ for $\phi_{jk} \in \left(-\frac{\pi}{4}+n\frac{\pi}{2}+m\pi,\frac{\pi}{4}+n\frac{\pi}{2}+m\pi\right)$, where $n=0,1$ and $m=0,1$. With this simplification, the QL still hosts a QI~\cite{Supplement}.

 According to the Kirchhoff's law~\cite{Imhof2018NatPhys,Lee2018ComPhys,Hofmann2019PRL,Ezawa2018PRBR,Jiang2015PRL,Luo2018Research,Luo2018arXiv}, we have
\begin{equation}
I_{pa}\left( \omega \right) =\sum_{q,b}J_{pa,qb}\left( \omega \right) V_{qb}\left(
\omega \right),
\end{equation}
where $I_{pa}(t)=I_{pa}\left(\omega\right)e^{i\omega t}$ [$V_{qb}(t)=V_{qb}\left(\omega\right)e^{i\omega t}]$ is the current (voltage) at node $a$ (b) in cell $p$ ($q$), and $\omega$ is the frequency of circuit.
The circuit Laplacian is $J_{pa,qb}\left( \omega \right) =i\omega H_{pa,qb}(\omega)$, where
\begin{equation}\label{circuit}
H_{pa,qb}(\omega)=C_{pa,qb}-\frac{1}{\omega^2}W_{pa,qb}.
\end{equation}
Here, $H$ is just the Hamiltonian of the tight-binding~(TB) model of the EC. $C_{pa,qb}$ is the capacitance between two nodes, and $W_{pa,qb}=L^{-1}_{pa,qb}$ is the inverse inducitivity between two nodes. For the diagonal components with $pa=qb$, we have $C_{pa,pa}=-C_{pa,g}-\sum_{q'b'}C_{pa,q'b'}$, and $W_{pa,pa}=-L^{-1}_{pa,g}-\sum_{q'b'}L^{-1}_{pa,q'b'}$. Subscript $g$ means the ground. In principle, the EC Hamiltonian in Eq.~\eqref{circuit} can be considered as the TB Hamiltonian of QL, expect for $D(\omega)$ \textit{i}.\textit{e}., $H(\omega)=H_\text{QL}(\omega)+D(\omega)$, and $D(\omega)$ is a diagonal matrix collecting all the diagonal elements of $H(\omega)$.
Now we discuss how to measure the CSs in the EC. As illustrated in Ref.~\cite{Imhof2018NatPhys}, the impedance between two nodes $Z_{pa,qb}(\omega) \equiv (V_{pa}-V_{qb})/I_0$ can be directly measured in experiments. Diagonalizing the $J$ matrix, we get $J_{pa,qb}(\omega)=\sum_l j_l(\omega) |\psi_l(pa)\rangle \langle\psi_l(qb) |$, where $j_l$ is the $l^{th}$ eigenvalue and $|\psi_l\rangle$ is the corresponding eigenvector. Then,
\begin{equation}
Z_{pa,qb}\left( \omega \right)  =\sum_{l}\frac{\left\vert \psi _{l}\left( pa\right) -\psi _{l}\left(
	qb\right) \right\vert ^2}{j_{l}\left( \omega \right) },
\end{equation}%
and $Z_{pa,qb}$ diverges whenever $j_{l}\left( \omega \right)=0$. The topological CS of $H_\text{QL}$ gives rise to $j_{l_0}=0$ when $D(\omega)=0$ at the resonance frequency $\omega_0$, so that $Z(\omega_0)$ shows a peak. $D(\omega_0)=0$ is guaranteed by a suitable choice of grounding~\cite{Supplement}. Meanwhile, we choose pairs of capacitors and inductors~($C_{0,1,2,3}$ and $L_{0,1,2,3}$) in the EC [Fig.~\ref{fig4}(b)] to satisfy $\omega_0=1/\sqrt{C_0L_0}=1/\sqrt{C_1L_1}=1/\sqrt{C_2L_2}=1/\sqrt{C_3L_3}$. Of course, the CS induced resonance peak can only be observed when the nodes are at the corner.

Figure~\ref{fig4}(c) shows the spectrum of the circuit Laplacian as a function of the normalized frequency $\omega/\omega_{0}$. Four degenerate isolated ZEMs inside the bulk spectral gap at $\omega_0$. In Fig.~\ref{fig4}(d), we plot the impedance between two nearest-neighbor nodes in one circuit cell located at the corner, the edge and in the bulk, respectively. We can see that the corner impedance shows an obvious resonance peak at $\omega_0$, indicating a quasicrystalline QI.

\emph{Conclusion.}---In summary, we have constructed quasicrystalline SOTIs by two distinct ways. One is the edge gapped quasicrystalline QSH insulator, and the other is the quasicrystalline QI. The quasicrystalline SOTIs exhibit some unusual characteristics. We also propose that the quasicrystalline QI can be simulated by an EC. Our work indicates that the concept of HOTIs is also valid in quasicrystalline systems, which enriches the HOTI family.

\emph{Acknowledgments.}---We would like to thank Xin Liu, Jin-Hua Sun, Yi Zhou, Rui Yu and Zhi-Hong Hang for helpful discussions. R.C. and D.-H.X. were supported by the NSFC (Grant No. 11704106). J.H.G. was supported by the NSFC (Grant No. 11534001 and 11874160). C.Z.C.  was funded by the NSFC (Grant No. 11974256) and the NSF of Jiangsu Province (Grant No. BK20190813). D.-H.X. also acknowledges the financial support of the Chutian Scholars Program in Hubei Province.

\emph{Note added.}---Recently, we became aware of a complementary study, which focuses on realizing a higher-order topological superconductor phase on an 8-fold symmetric patch of the AB tiling~\cite{DanielVarjas2019arXiv}.

\end{document}


\title{Supplementary Materials to ``Higher-Order Topological Insulators in Quasicrystals''}
\author{Rui Chen}
\affiliation{Department of Physics, Hubei University, Wuhan 430062, China}
\author{Chui-Zhen Chen}
\affiliation{Institute for Advanced Study and School of Physical Science and Technology, Soochow University, Suzhou 215006, China}
\author{Jin-Hua Gao}
\thanks{jinhua@hust.edu.cn}
\affiliation{School of Physics and Wuhan National High Magnetic Field Center, Huazhong University of Science and Technology, Wuhan 430074, China}
\author{Bin Zhou}\thanks{binzhou@hubu.edu.cn}
\author{Dong-Hui Xu} \thanks{donghuixu@hubu.edu.cn}
\affiliation{Department of Physics, Hubei University, Wuhan 430062, China}

\maketitle

\renewcommand{\thefigure}{S\arabic{figure}}
\renewcommand{\thetable}{S\arabic{table}}
\renewcommand{\theequation}{S\arabic{equation}}
\renewcommand{\figurename}{Fig.}
\renewcommand{\tablename}{Tab.}

In this Supplementary Material, we first present the topological classification and stability of the second-order topological insulator~(SOTI) models. We then show the results of the mass term induced SOTIs on the Ammann-Beenker~(AB) tiling quasicrystalline lattice~(QL) with other boundaries and on the Penrose tiling QL. We also display the results of quasicrystalline quadrupole insulator~(QI) on the Penrose tiling QL, and provide more details about how to design the electrical circuit~(EC) to engineer the quasicrystalline QI.

\section{I. Mass term induced SOTIs in the AB tiling quasicrystal: Symmetries, topological invariant and stability}

In this Section, we investigate the topological properties of the mass term induced SOTIs in quasicrystals by symmetries, perturbations and topological invariant.
Table~\ref{tab11} illustrates the symmetries of the model $H_\text{I}=H_\text{QSH}+H_\text{m}(\eta)$ [Eq.~(1) and Eq.~(2) in the main text] with ($g\neq 0$) and without ($g=0$) the mass term $H_\text{m}$ for the square- or octagon-shaped AB tiling quasicrystals. When $g=0$, the Hamiltonian $H_\text{I}=H_\text{QSH}$ respects time-reversal symmetry~(TRS)~$T=U_TK$, particle-hole symmetry~(PHS)~$C=U_CK$ and chiral symmetry~$S=TC$, where unitary matrices $U_{T,C}$ satisfy $U_{T}U_{T}^{\ast }=-1$, $U_{C}U_{C}^{\ast }=1$, and $K$ is complex conjugation. Therefore, $H_\text{QSH}$ belongs to the symmetry class DIII and can describe the quantum spin Hall state or a time-reversal symmetric topological superconductor. When $g\neq 0$, in addition to breaking TRS, the mass term $H_\text{m}(\eta=2)$ breaks rotation symmetry $C_4$ and mirror symmetry $m_z$ but preserves the combined symmetry $C_{4}m_z$, while $H_\text{m}(\eta=4)$ breaks symmetries $C_8$ and $m_z$ but preserves the combined symmetry $C_{8}m_z$. We found that the zero-energy corner states~(CSs) in the square- and octagon-shaped AB tiling quasicrystals are protected by the combined symmetries  $C_{4,8}m_z$ and PHS $C$.
More details about symmetries as shown in Table~\ref{tab11}.
\renewcommand\arraystretch{1.2}
\begin{table*}[hbtp]
\begin{tabular}{|c|c|cc|c|c|}
\hline
&  & \multicolumn{2}{c|}{Square} & \multicolumn{2}{c|}{Octagon} \\
\hline
&  & $g=0$ & \multicolumn{1}{|c|}{$g\neq 0,\eta =2$} & $g=0$ &
\multicolumn{1}{c|}{$g\neq 0,\eta =4$} \\ \hline
$T=U_{T}K$ & $THT^{-1}=H$ & $\checkmark $ & \multicolumn{1}{|c|}{$\times $}
& $\checkmark $ & $\times $ \\ \hline
$C=U_{C}K$ & $CHC^{-1}=-H$ & $\checkmark $ & \multicolumn{1}{|c|}{$%
\checkmark $} & $\checkmark $ & $\checkmark $ \\ \hline
$S=TC$ & $SHS^{-1}=-H$ & $\checkmark $ & \multicolumn{1}{|c|}{$\times $} & $%
\checkmark $ & $\times $ \\ \hline
$m_{x}=s_{1}\tau _{0}\mathcal{M}_{x}$ & $m_{x}Hm_{x}^{-1}=H$ & $\checkmark $
& \multicolumn{1}{|c|}{$\checkmark $} & $\checkmark $ & $\checkmark $ \\
\hline
$m_{y}=s_{2}\tau _{3}\mathcal{M}_{y}$ & $m_{y}Hm_{y}^{-1}=H$ & $\checkmark $
& \multicolumn{1}{|c|}{$\checkmark $} & $\checkmark $ & $\checkmark $ \\
\hline
$m_{z}=s_{3}\tau _{0}$ & $m_{z}Hm_{z}^{-1}=H$ & $\checkmark $ &
\multicolumn{1}{|c|}{$\times $} & $\checkmark $ & $\times $ \\ \hline
$C_{4}=e^{-i\frac{\pi }{4}s_{3}\tau _{3}}\mathcal{R}_{4}$ & $%
C_{4}HC_{4}^{-1}=H$ & $\checkmark $ & \multicolumn{1}{|c|}{$\times $} & $%
\checkmark $ & $\checkmark $ \\ \hline
$C_{8}=e^{-i\frac{\pi }{8}s_{3}\tau _{3}}\mathcal{R}_{8}$ & $%
C_{8}HC_{8}^{-1}=H$ & $ $ & \multicolumn{1}{|c|}{$ $} & $%
\checkmark $ & $\times $ \\ \hline
$C_{4}T$ & $C_{4}TH\left( C_{4}T\right) ^{-1}=H$ & $\checkmark $ &
\multicolumn{1}{|c|}{$\checkmark $} & $\checkmark $ & $\times $ \\ \hline
$C_{8}T$ & $C_{8}TH\left( C_{8}T\right) ^{-1}=H$ & $ $ &
\multicolumn{1}{|c|}{$ $} & $\checkmark $ & $\checkmark $ \\ \hline
$C_{4}m_{x}$ & $C_{4}m_{x}H\left( C_{4}m_{x}\right) ^{-1}=H$ & $\checkmark $
& \multicolumn{1}{|c|}{$\times $} & $\checkmark $ & $\checkmark $ \\ \hline
$C_{4}m_{y}$ & $C_{4}m_{y}H\left( C_{4}m_{y}\right) ^{-1}=H$ & $\checkmark $
& \multicolumn{1}{|c|}{$\times $} & $\checkmark $ & $\checkmark $ \\ \hline
$C_{4}m_{z}$ & $C_{4}m_{z}H\left( C_{4}m_{z}\right) ^{-1}=H$ & $\checkmark $
& \multicolumn{1}{|c|}{$\checkmark $} & $\checkmark $ & $\times $ \\ \hline
$C_{8}m_{x}$ & $C_{8}m_{x}H\left( C_{8}m_{x}\right) ^{-1}=H$ & $ $ &
\multicolumn{1}{|c|}{$ $} & $\checkmark $ & $\times $ \\ \hline
$C_{8}m_{y}$ & $C_{8}m_{y}H\left( C_{8}m_{y}\right) ^{-1}=H$ & $ $ &
\multicolumn{1}{|c|}{$ $} & $\checkmark $ & $\times $ \\ \hline
$C_{8}m_{z}$ & $C_{8}m_{z}H\left( C_{8}m_{z}\right) ^{-1}=H$ & $ $ &
\multicolumn{1}{|c|}{$ $} & $\checkmark $ & $\checkmark $ \\ \hline
\end{tabular}%
\caption{Symmetries of the SOTI model $H_\text{I}$ with ($g\neq 0$) and without ($g=0$) the mass term
$H_\text{m}(\eta)$ for AB tiling square and octagon geometries. Here unitary matrices $U_{T}=is_{2}\protect\tau %
_{0}$ and $U_{C}=s_{3}\protect\tau _{1}$. $\mathcal{R}_{4,8}$ are orthogonal matrices permuting
the sites of the tiling to rotate the whole system by $\protect%
\pi/2$ and $\protect\pi/4$, while $\mathcal{M}_{x,y}$ are orthogonal matrices
permuting the sites of the tiling to flip the whole system vertically and
horizontally. }
\label{tab11}
\end{table*}

\begin{figure*}[pthb]
	\centering
	\includegraphics[width=15cm]{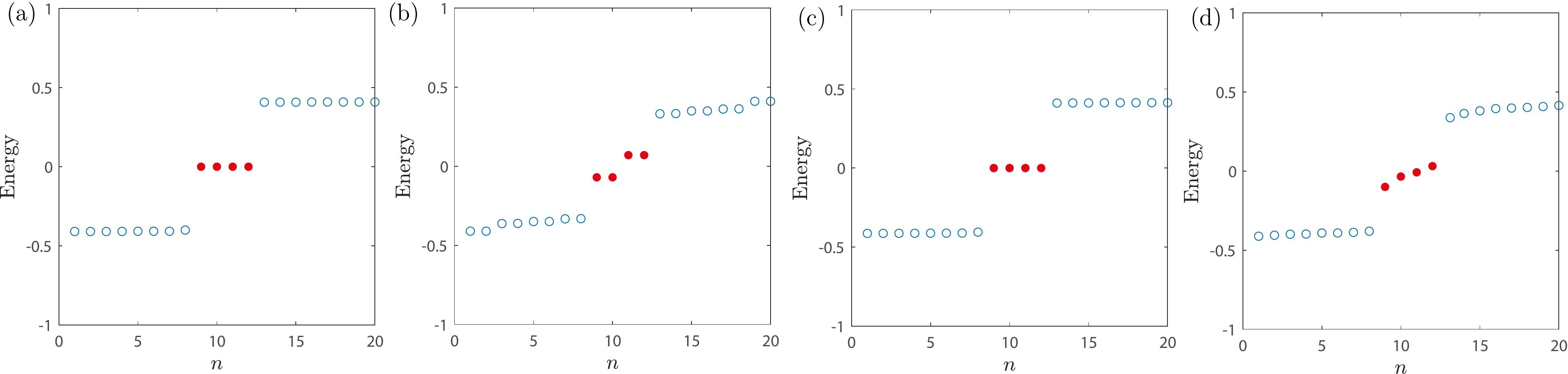}
	\caption{ Energy spectra on the AB tiling QL with the square boundary for (a) $H_{\text{I}}+\Delta H_1$,
(b) $H_{\text{I}}+\Delta H_2$,
(c) $H_{\text{I}}+\Delta H_3$,  and
(d) $H_{\text{I}}+\Delta H_4$. Here we set the parameters $U^{1,2}=W_{1,2}=0.1$, $g=1$, $M=-1$, and $\eta=2$. The lattice sites are 4061. }
	\label{fig4}
\end{figure*}

\begin{figure*}[pthb]
	\centering
	\includegraphics[width=15cm]{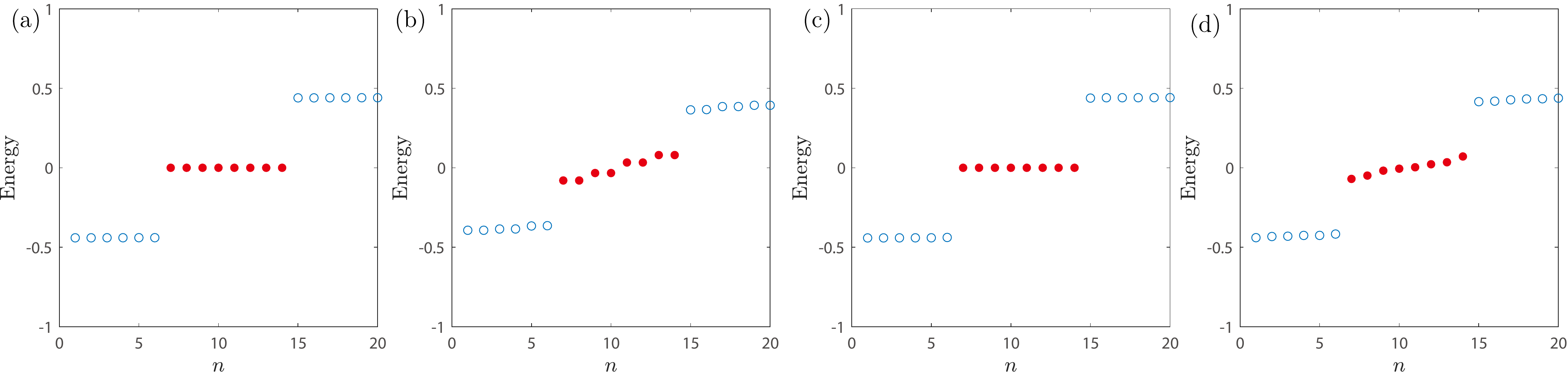}
	\caption{ Energy spectra on the AB tiling QL with the octagon boundary for (a) $H_{\text{I}}+\Delta H_1$,
(b) $H_{\text{I}}+\Delta H_2$,
(c) $H_{\text{I}}+\Delta H_3$, and
(d) $H_{\text{I}}+\Delta H_4$. Here we set the parameters $U^{1,2}=W_{1,2}=0.1$, $g=1$, $M=-1$, and $\eta=4$. The lattice sites are 4713. }
	\label{fig4_1}
\end{figure*}

To investigate the robustness of the zero-energy CSs, we consider the following two kinds of symmetry breaking perturbations
\begin{align}
\Delta H_1=\sum_{j} U^{1}c_{j}^{\dag }s _{3}\tau _{3} c_{j},\\
\Delta H_2=\sum_{j} U^{2}c_{j}^{\dag }s _{0}\tau _{1} c_{j},
\end{align}
where $U^{1,2}$ are uniform on-site potentials, $s_{1,2,3}$ and $\tau_{1,2,3}$ are the Pauli matrices acting on the spin and orbital degrees of freedom. $\Delta H_1$ preserves $C_{4,8}m_z$ and PHS $C$, while $\Delta H_2$ breaks all these symmetries. The energy spectrum of $H_\text{I}$ with $\eta=2$ under $\Delta H_1$ in the AB tiling quasicrystal square is displayed in Fig.~\ref{fig4}(a). We can see that the zero-energy CSs remain stable, this is because the symmetries that protect the CSs are not broken. Figure~\ref{fig4}(b) shows the results with the perturbation $\Delta H_2$. Apparently, the four zero-energy CSs are gapped out as $\Delta H_2$ breaks the symmetries provide the topological protection for CSs.
As shown in Figs.~\ref{fig4_1}(a) and~\ref{fig4_1}(b), the spectra of $H_\text{I}$ for an octagon geometry with $\eta=4$ under perturbations $\Delta H_{1,2}$ are plotted. The eight zero-energy CSs in the quasicrystal octagon are stable in the presence of $\Delta H_1$ but they are gapped out by $\Delta H_2$ due to losing the symmetry-protected topology.

\begin{figure}[hbtp]
	\centering
	\includegraphics[width=8cm]{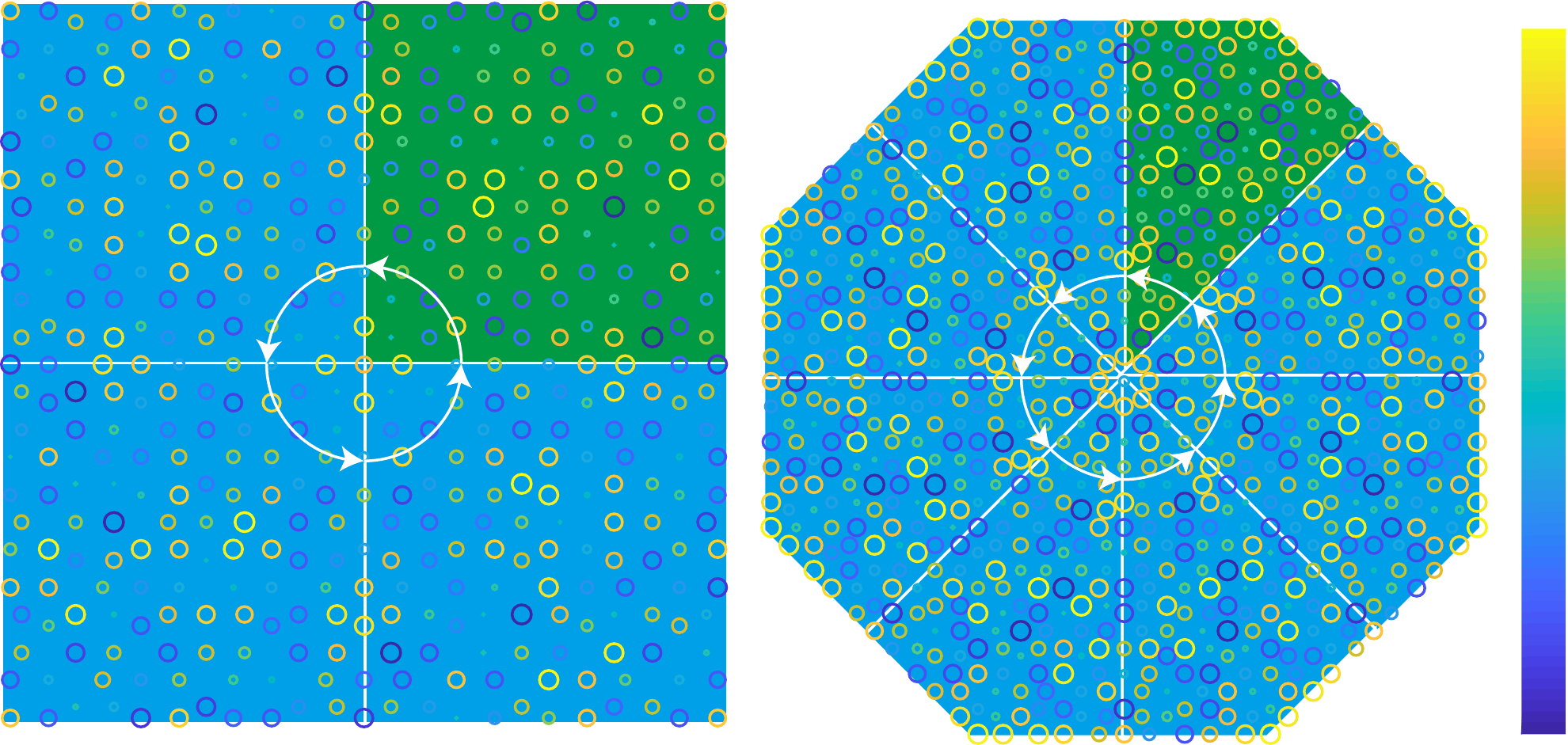}
	\caption{Schematic illustration of the $C_4$- and $C_8$-invariant distribution of the random potential $U_j^{1}$. The size and color of the circles label the magnitude of $U_j^{1}$.}
	\label{fig3}
\end{figure}

Furthermore, we can replace the uniform potentials $U^{1,2}$ by a random potential to study the disorder effect. Then $\Delta H_{1,2}$ become
\begin{align}
\Delta H_3=\sum_{j} U^{1}_{j}c_{j}^{\dag }s _{3}\tau _{3} c_{j},\\
\Delta H_4=\sum_{j} U^{2}_{j}c_{j}^{\dag }s _{0}\tau _{1} c_{j}.
\end{align}
where $U_{j}^{1,2}$ are uniformly distributed within the range $\left(-W_{1,2},W_{1,2}\right)$ with the disorder strength $W_{1,2}$. To maintain the rotation symmetries, we first generate $U_j^1$ in the green region, then $U_j^1$ in the cyan regions is obtained by rotating by an angle of $\pi/2$ or $\pi/4$ as shown in Fig.~\ref{fig3}. The influences of disorder are similar. As long as the disorder preserves $C_{4,8}m_z$ and $C$, the corner states are stable~[Figs.~\ref{fig4}(c) and ~\ref{fig4_1}(c)]. Otherwise, they are gapped by the disorder~[Figs.~\ref{fig4}(d) and ~\ref{fig4_1}(d)].


By the symmetry analysis, we know that the SOTIs on the AB tiling QL are protected by $C_{4,8}m_z$ and PHS $C$. To characterize the bulk topology of the SOTIs, we employ a method developed in Ref.~\cite{DanielVarjas2019arXiv} to calculate the bulk topological invariant of the Hamiltonian $H_\text{I}$. First, we introduce a momentum-dependent effective Green's function by projecting the single-particle Green's function onto plane-wave states
\begin{equation}
G_{\text{eff}}(k)=\left\langle k,a\right\vert G\left\vert k,b\right\rangle ,
\end{equation}%
where $\left\vert k,a\right\rangle $ is a normalized plane-wave state with
nonzero amplitude only in the local orbital $a$, and $G=\lim_{\delta
\rightarrow 0}\left( H_{\text{I}}+i\delta \right) $ is the Green's function of
the whole Hamiltonian at zero energy. For $\eta=2$, the SOTI is protected by $C_4m_z$ and $C$ symmetries, thus the effective Hamiltonian $H_{\text{eff}}=G_{\text{eff}}^{-1}$ satisfies
\begin{align}
H_{\text{eff}}^{S}\left( k\right) & =U_{C_{4}m_{z}}H_{\text{eff}}^{S}\left( R_{C_{4}m_{z}}^{-1}k\right)
U_{C_{4}m_{z}}^{-1},  \\
H_{\text{eff}}^{S}\left( k\right) & =-U_{C}H_{\text{eff}}^{S}\left( -k\right) ^{\ast
}U_{C}^{-1},
\end{align}
while for $\eta=4$, the SOTI is protected by $C_8m_z$ and $C$ symmetries, the effective Hamiltonian obeys
\begin{align}
H_{\text{eff}}^{O}\left( k\right) & =U_{C_{8}m_{z}}H_{\text{eff}}^{O}\left( R_{C_{8}m_{z}}^{-1}k\right)
U_{C_{8}m_{z}}^{-1},  \\
H_{\text{eff}}^{O}\left( k\right) & =-U_{C}H_{\text{eff}}^{O}\left( -k\right) ^{\ast
}U_{C}^{-1},
\end{align}
where $U_{C_4 m_z}$, $U_{C_8 m_z}$, and $U_C$ are the unitary actions of the $C_4 m_z$, $C_8 m_z$ and PHS $C$ operators, and $R_{C_4 m_z}$ and $R_{C_8 m_z}$ rotate the momenta by $\pi/2$ and $\pi/4$, respectively.
\begin{figure}[hbtp]
	\centering
	\includegraphics[width=16cm]{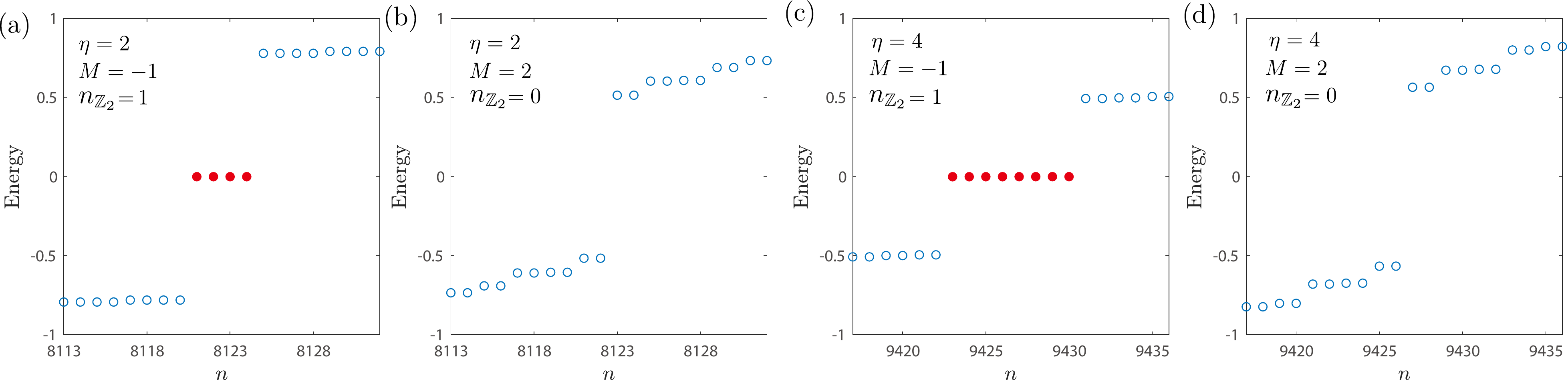}
	\caption{Energy spectra on the AB tiling QL with the square and octagon geometries. (a) $H_{\text{I}}^{S}$ with $M=-1$ and $n_{\mathbb{Z}_2}=1$, (b) $H_{\text{I}}^{S}$ with $M=2$ and $n_{\mathbb{Z}_2}=0$, (c) $H_{\text{I}}^{O}$ with $M=-1$ and $n_{\mathbb{Z}_2}=1$, and (d) $H_{\text{I}}^{O}$ with $M=2$ and $n_{\mathbb{Z}_2}=0$. Here we take the parameter $g=1$, $\eta=2$ in (a) and (b), and $\eta=4$ in (c) and (d). The lattice sites are 4061 for (a, b) and 4713 for (c, d). Only 20 eigenvalues near the zero energy are illustrated.}
	\label{fig1}
\end{figure}

Since the two-dimensional (2D) AB tiling QL can be generated from a 4D cubic lattice, we consider the plane-wave states defined in the corresponding 4D reciprocal space. Plane-wave states in the 4D Brillouin zone form an over-complete basis for the QL, however, topological invariants are only determined by the Pfaffians of the effective Hamiltonian at the $C_4$- or $C_8$-invariant momenta~\cite{DanielVarjas2019arXiv}. $C_8$-invariant momenta include $\Gamma=\left(0,0,0,0\right)$ and $\Pi=\left(\pi,\pi,\pi,\pi\right)$, while the $C_4$-invariant momenta consist of $\Gamma$, $\Pi$ and two additional high symmetry points $X=\left(0,\pi,0,\pi\right)$ and $X'=\left(\pi,0,\pi,0\right)$. When $\eta=4$, we obtain four invariants $\nu_{1,3,5,7}^{\Gamma}$ by calculating the Pfaffians of the effective Hamiltonian $H_{\text{eff}}^{O}$ in the eigen-subspaces of $C_{8}m_z$ and $C$ at $\Gamma$ point. The effective Hamiltonian $H_{\text{eff}}^{O}$ at point $\Pi$ can be easily obtained by flipping the sign of every next-neighbor-hopping in the original model. Similarly, we can obtain invariants $\nu_{1,3,5,7}^{\Pi}$ at point $\Pi$. The topological $\mathbb{Z}_2$ invariant $n_{\mathbb{Z}_2}$ characterizing the SOTI is given by $\left(-1\right)^{n_{\mathbb{Z}_2}}=\nu_{1}\nu_{3}$ with $\nu_{i}=\nu_{i}^{k=0}/\nu_{i}^{k=\Pi}$, where $i=1,3$. Here, we have used the fact that there are only two independent invariants, i.e., $\nu_1=\nu_7$ and $\nu_3=\nu_5$. When $\eta=2$, it is trickier to deal with the effective Hamiltonian at points $X$ and $X'$. To compute the topological invariant, we only keep hoppings to the next-nearest-neighbors as we can flip half of the next-nearest-neighbor hoppings to get $H_{\text{eff}}^{S}$ at $X$, and then flip the other half of the next-nearest-neighbor hoppings at point $X'$. It is found that the invariants at $X$ and $X'$ points are equal to each other and thus do not contribute to the $\mathbb{Z}_2$ invariant. Finally, the topological $\mathbb{Z}_2$ invariant $n_{\mathbb{Z}_2}$ in this case is given by $\left(-1\right)^{n_{\mathbb{Z}_2}}=\nu_{1}$ with $\nu_{1}=\nu_{1}^{k=0}/\nu_{1}^{k=\Pi}$.
\begin{figure}[hbtp]
	\centering
	\includegraphics[width=12cm]{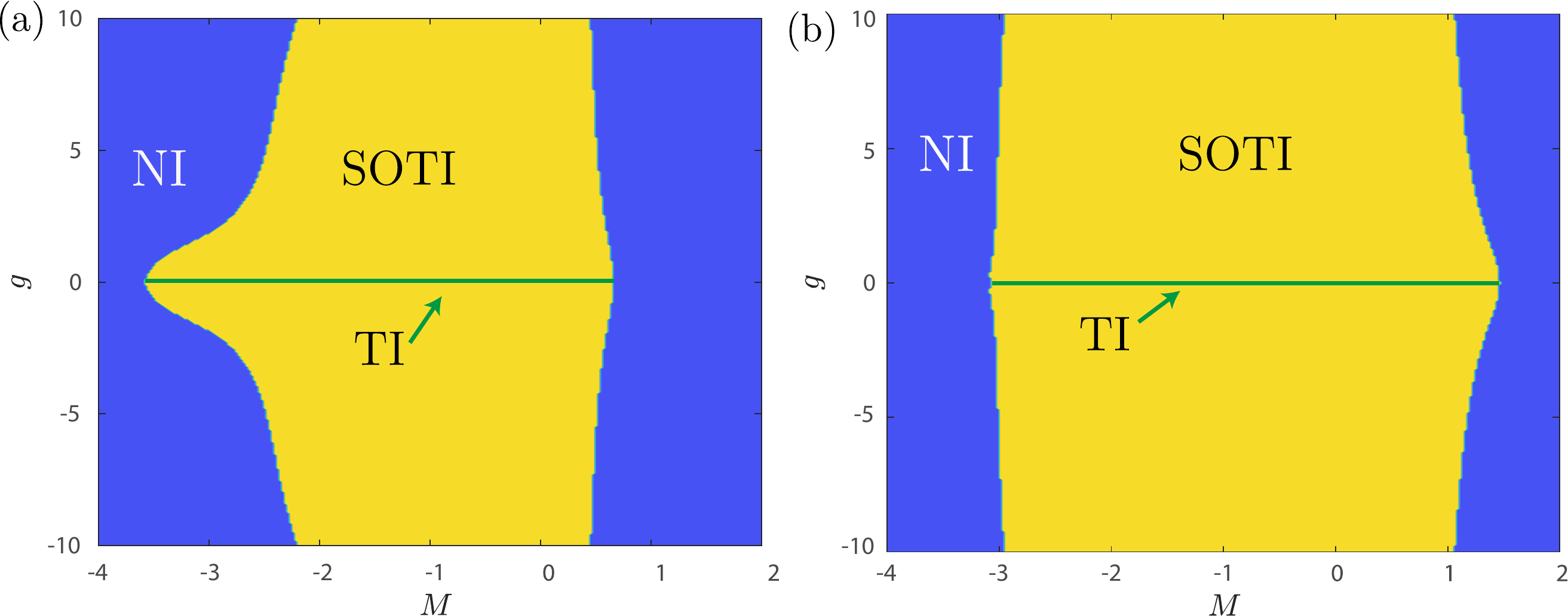}
	\caption{(a) Phase diagram of the $\mathbb{Z}_2$ invariant in the $\left(M,g\right)$ space with the mass term $H_\text{m}(\eta=2)$. The yellow and blue regions correspond to the SOTI with $n_{\mathbb{Z}_2}=1$ and the normal insulator~(NI) with $n_{\mathbb{Z}_2}=0$, respectively. The green line denotes the first-order topological insulator~(TI) phase since $g=0$ along the line. The results for $H_\text{m}(\eta=4)$ are shown in (b).}
	\label{fig2}
\end{figure}

Figure~\ref{fig1} shows the energy spectra versus the topological invariant $n_{\mathbb{Z}_2}=1$ when $g=1$ for (a) $\eta=2$, $M=-1$, (b) $\eta=2$, $M=2$, (c) $\eta=4$, $M=-1$ and (d) $\eta=4$, $M=2$. We can see that the zero-energy CSs appear when $n_{\mathbb{Z}_2}=1$, and there is no zero-energy CS when $n_{\mathbb{Z}_2}=0$. Therefore, this topological invariant can be used to characterize HOTIs in quasicrystals. Moreover, in Fig.~\ref{fig2} we illustrate the phase diagrams of $H_\text{I}$ on the AB tiling QL with $\eta=2, 4$ in the $\left(M,g\right)$ space obtained by computing the $\mathbb{Z}_2$ invariant. One should notice that the energy gaps of the edge states close for $g=0$ [the green lines in Figs.~\ref{fig2}(a) and \ref{fig2}(b)], which corresponds to the first-order topological insulator.

\section{II. Quasicrystalline QI: Symmetries, topological invariant and stability}
\begin{figure}[pthb]
	\centering
	\includegraphics[width=12cm]{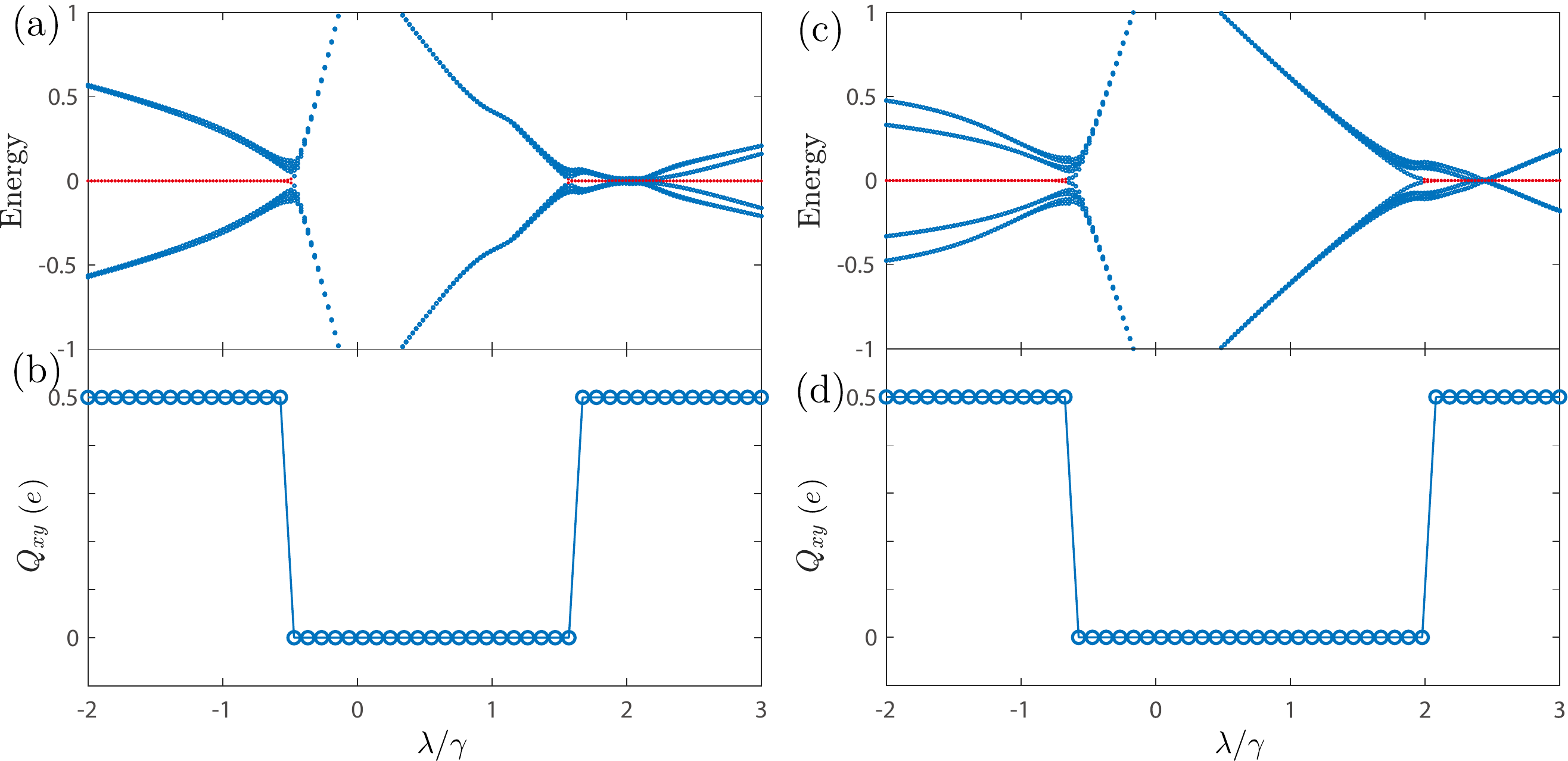}
	\caption{ Energy spectra and the quadrupole moment $Q_{xy}$ of QI models as a function of the parameter $\lambda/\gamma$ in the AB tiling quasicrystal square. (a) and (b) Model $H_{\text{QI}}$. (c) and (d) Model $H_{\text{QI}}'$. Here only the 20 energy states close to the zero energy are illustrated in (a) and (c), and the number of cells is 4061. The ZEMs are marked by red dots.}
	\label{fig5}
\end{figure}

In this section, we explore the symmetries, topological invariant and stability of the QI models on the AB tiling QL. In Tab.~\ref{tab12}, we list the symmetries of the QI models $H_{\text{QI}}$ and $H_{\text{QI}}'$, where $H_{\text{QI}}'$ denotes the QI model $H_{\text{QI}}$ with a simplified $T(\protect\phi_{jk})$. Both $H_{\text{QI}}$ and $H_{\text{QI}}'$ preserve TRS and PHS, meanwhile $U_{T}^{\ast }U_{T}=U_{C}^{\ast }U_{C}=1$, therefore the quasicrystalline QI models belong to the class BDI. Moreover, $H_{\text{QI}}$ respects two mirror symmetries $m_{x,y}$ and $C_4$ symmetry, while for $H_{\text{QI}}'$ the mirror symmetries $m_{x,y}$ are broken, but it still preserves the $C_4$ symmetry. The quadrupole moment $Q_{xy}$ is quantized to $0$ or $e/2$ when there exists mirror symmetries $m_{x}$ and $m_y$ or $C_4$ symmetry~\cite{Wheeler2018arXiv,Kang2018arXiv}. The quantized quadrupole moment $Q_{xy}$ can be used to characterize the topology of $H_{\text{QI}}$ and $H_{\text{QI}}'$~\cite{Wheeler2018arXiv,Kang2018arXiv}. We illustrate the quadrupole moment $Q_{xy}$ of $H_{\text{QI}}$ and $H_{\text{QI}}'$ in Figs.~\ref{fig5}(b) and \ref{fig5}(d), respectively. The corresponding energy spectra are also plotted in Fig.~\ref{fig5}. We found that zero-energy modes~(ZEMs) localized at the sample corners appear when $Q_{xy}=e/2$, which reveals the topological origin of the CSs.
\renewcommand\arraystretch{1.2}
\begin{table}[pthb]
	\begin{tabular}{|c|c|c|c|}
		\hline
		&  & $H_{\text{QI}}$ & $H_{\text{QI}}'$ \\ \hline
		$T=U_{T}K$ & $THT^{-1}=H$ & $\checkmark $ & $\checkmark $ \\ \hline
		$C=U_{C}K$ & $CHC^{-1}=-H$ & $\checkmark $ & $\checkmark $ \\ \hline
		$S=TC$ & $SHS^{-1}=-H$ & $\checkmark $ & $\checkmark $ \\ \hline
		$m_{x}=\tau _{1}\tau _{3}\mathcal{M}_{x}$ & $m_{x}Hm_{x}^{-1}=H$ & $%
		\checkmark $ & $\times $ \\ \hline
		$m_{y}=\tau _{1}\tau _{1}\mathcal{M}_{y}$ & $m_{y}Hm_{y}^{-1}=H$ & $%
		\checkmark $ & $\times $ \\ \hline
		$C_{4}=U_{C_{4}}\mathcal{R}_{4}$ & $C_{4}HC_{4}^{-1}=H$ & $\checkmark $ & $%
		\checkmark $ \\ \hline
	\end{tabular}%
	\ \ \ \ \ \
	\caption{Symmetries of the SOTI models $H_\text{QI}$ and $H_\text{QI}'$ on the AB tiling quasicrystal square. Here $U_{T}=\tau _{0}\tau _{0}$, $U_{C}=\tau _{3}\tau _{0}$, and $U_{C_{4}}=%
		\begin{pmatrix}
		0 & i\tau _{2} \\
		\tau _{0} & 0%
		\end{pmatrix}%
		$. }
	\label{tab12}
\end{table}

\begin{figure*}[pthb]
	\centering
	\includegraphics[width=16cm]{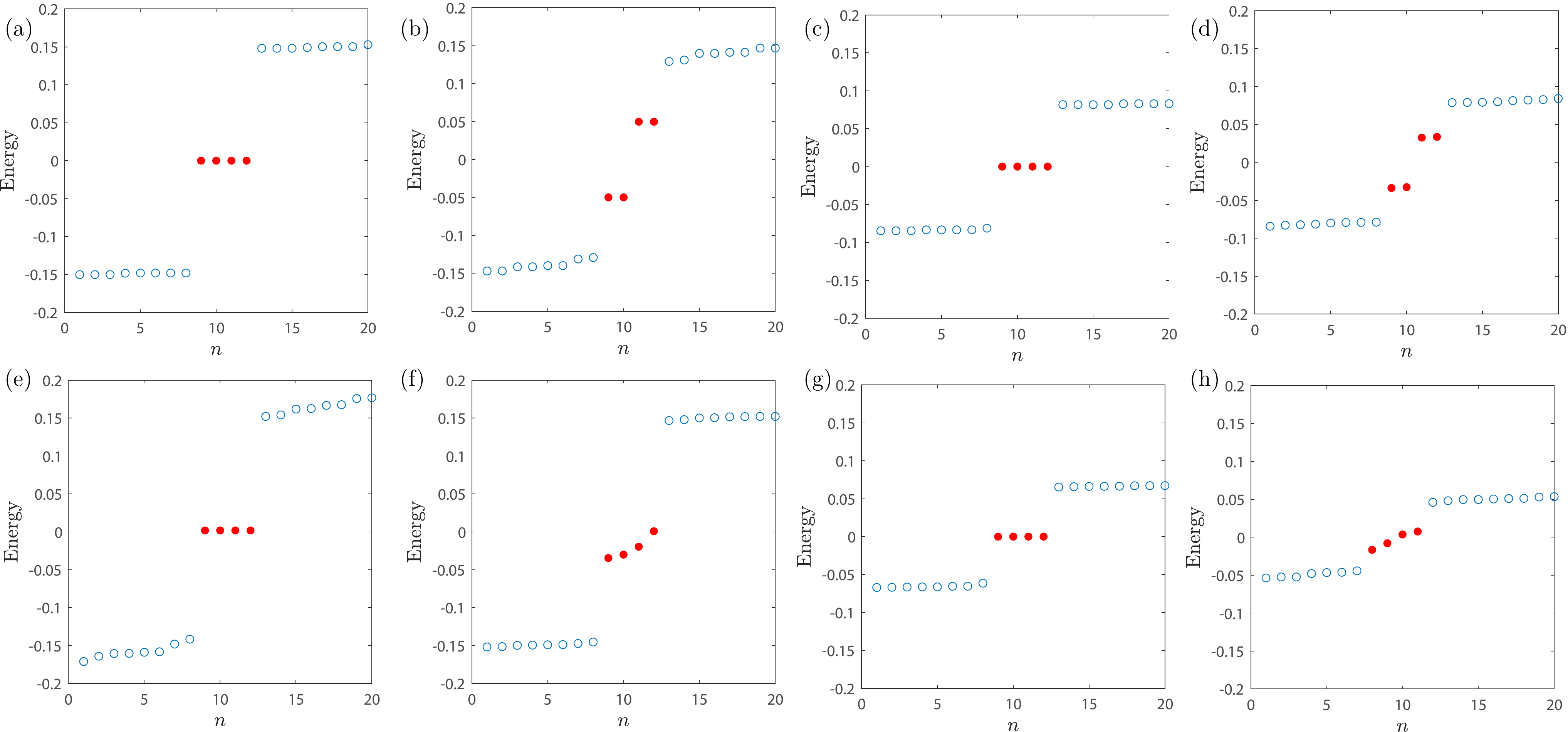}
	\caption{The spectra the QI models on the AB tiling QL for
(a) $H_{\text{QI}}$ with $\Delta H_1'$.
(b) $H_{\text{QI}}$ with $\Delta H_2'$.
(c) $H_{\text{QI}}'$ with $\Delta H_1$.
(d) $H_{\text{QI}'}$ with $\Delta H_2'$.
(e) $H_{\text{QI}}$ with $\Delta H_3'$.
(f) $H_{\text{QI}}$ with $\Delta H_4'$.
(g) $H_{\text{QI}}'$ with $\Delta H_3'$.
(h) $H_{\text{QI}'}$ with $\Delta H_4'$.
 Here we take $U^{1',2'}=W_{1',2'}=0.05$. The number of cells is 4061. }
	\label{fig6}
\end{figure*}

To investigate the stability of the CSs in $H_{\text{QI}}$ and $H_{\text{QI}}'$, we introduce the following two perturbation terms
 \begin{align}
\Delta H_1'&=\sum_{j} U^{1'}c_{j}^{\dag }\tau _{0}\tau _{2} c_{j},\\
\Delta H_2'&=\sum_{j} U^{2'}c_{j}^{\dag }\tau _{0}\tau _{3} c_{j}.
\end{align}
 As shown in Figs.~\ref{fig6}(a) and \ref{fig6}(c), the four zero-energy CSs remain stable in the presence of $\Delta H_1'$ since $C_4$ symmetry is preserved. When introducing $\Delta H_2'$, the zero-energy CSs are gapped out [Figs.~\ref{fig6}(b) and \ref{fig6}(d)] due to both $C_4$ and $m_{x,y}$ symmetries are broken by $\Delta H_2'$.

We can study the disorder effect by considering the following random potentials
 \begin{align}
\Delta H_3'&=\sum_{j} U^{1'}_{j}c_{j}^{\dag }\tau _{0}\tau _{2} c_{j},\\
\Delta H_4'&=\sum_{j} U^{2'}_{j}c_{j}^{\dag }\tau _{0}\tau _{3} c_{j}.
\end{align}
Similar to the previous section, we require $\Delta H_3'$ to satisfy $C_4$ symmetry when generating $U^{1'}_j$, while we don't restrict $U^{2'}_j$ to obey $C_4$ symmetry. As shown in Figs.~\ref{fig6}(e) and \ref{fig6}(g), the four zero-energy CSs remain stable in the presence of $\Delta H_3'$. When introducing $\Delta H_4'$, symmetries $C_4$ and $m_{x,y}$ are broken, thus the zero-energy CSs are gapped out [Figs.~\ref{fig6}(f) and \ref{fig6}(h)].

\section{III. Mass term induced SOTI on the AB tiling QL with other boundaries}

\begin{figure*}[pthb]
	\centering
	\includegraphics[width=16cm]{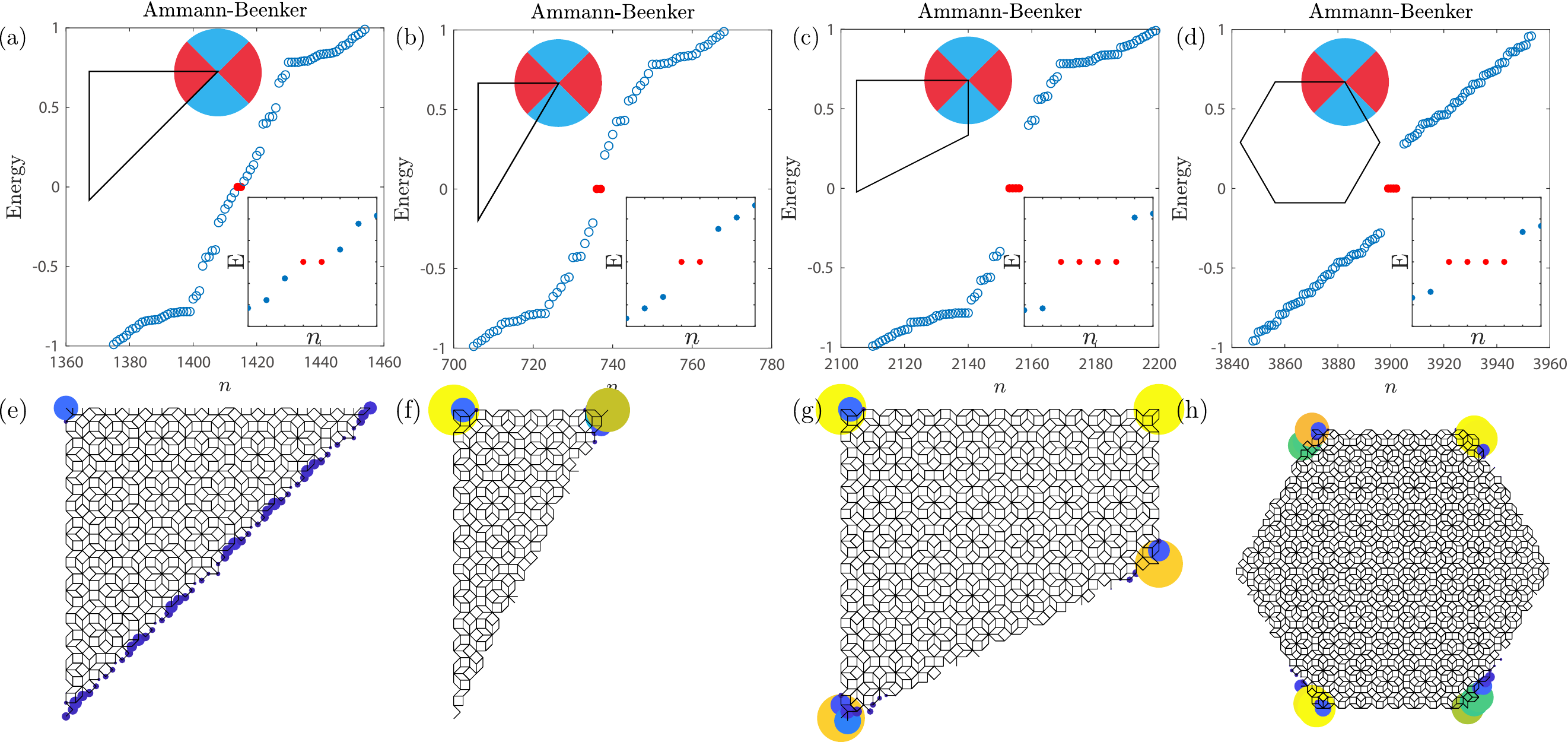}
	\caption{(a-d) The spectra of the SOTI model $H_\text{I}$ with $\protect\eta%
		=2$ on the AB tiling QL with the isosceles right triangle, right triangle,
		trapezoid, and hexagon boundaries. The insets denote the zoomed-in section
		of ZEMs marked as the red dots. (e-h) The wavefunction probability of ZEMs
		in (a-d).}
	\label{fig7}
\end{figure*}

In this Section, we give the results of SOTI model $H_\text{I}$ with $\eta=2$ on the AB
tiling QL under different boundary geometries. The energy spectra and
wavefunction probability of the ZEMs are displayed in
Fig.~\ref{fig7}. The blue and red regions in Figs.~\ref{fig7}(a), \ref%
{fig1}(b), \ref{fig7}(c) and \ref{fig7}(d) denote the two different
regions of the edge orientation that have opposite sign of the Wilson mass.
According to the Jackiw-Rebbi mechanism~\cite{Jackiw1976PRD}, a topological
ZEM appears when a mass domain wall forms. Then, we can determine the
spatial profile of ZEMs by observing the edge orientation, which is in
accordance with the numerical results. For example, there exist two ZEMs for
the isosceles right triangle [Fig.~\ref{fig7}(a)], one is the localized
CS and the other is the extended state spreading along one
edge of the triangle~[Fig.~\ref{fig7}(e)]. The CS can be explained that
the two adjacent edges that make the corner have the effective Wilson mass
with opposite sign since they are in the red region and the blue region,
receptively. This extended state appears due to the effective Wilson mass
for the edge is zero. For the right triangle in Fig.~\ref{fig7}(b), two
CSs are localized at the top two corners as shown in Fig.~\ref{fig7}(f).
However, CS is absent at the bottom corner. That is because the two adjacent
edges make the bottom corner all lie in the blue region. For the trapezoid
[Fig.~\ref{fig7}(c)] and hexagon [Fig.~\ref{fig7}(d)], our numerical
calculations show four CSs localized at the four corners as shown in Figs.~%
\ref{fig7}(g), and \ref{fig7}(h). Similarly, we can determine which of the
corners can host a CS by observing the edge orientation.

\section{IV. Mass term induced SOTI on the Penrose tiling QLs}

\begin{figure*}[pthb]
	\centering
	\includegraphics[width=16cm]{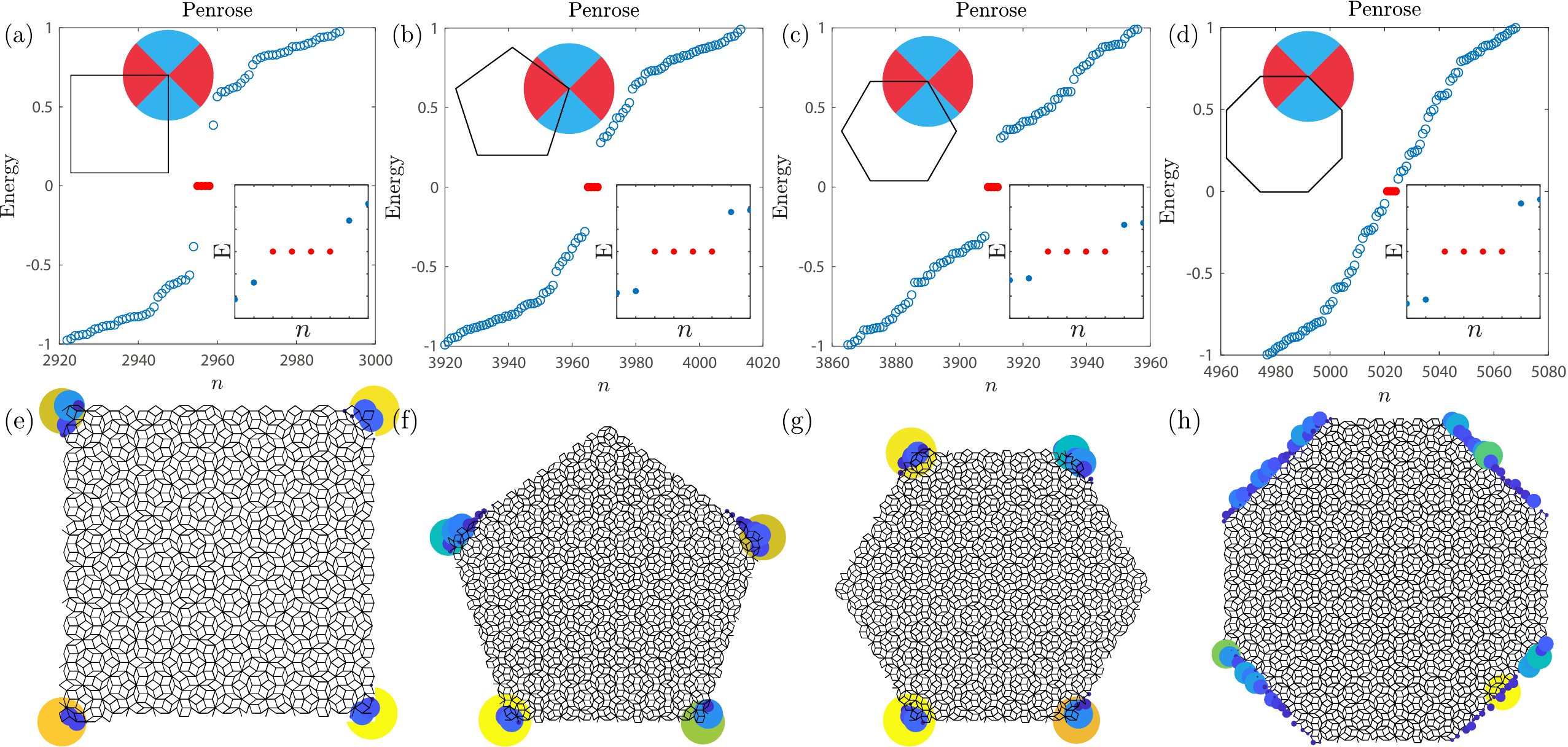}
	\caption{(a-d) The spectra of the SOTI model $H_\text{I}$ with $\protect\eta%
		=2$ on the Penrose tiling QL with the square, pentagon, hexagon and octagon
		boundaries. The insets denote the zoomed-in section of ZEMs marked as the
		red dots. (e-h) The wavefunction probability of ZEMs in (a-d).}
	\label{fig8}
\end{figure*}

The mass term induced SOTI can also be realized on the Penrose tiling QL.
Figure~\ref{fig8} shows the energy spectra and the wavefunction
probability of ZEMs for the SOTI model $H_\text{I}$ with $\eta=2$ on the
Penrose tiling QL with the square [Figs.~\ref{fig8}(a) and \ref{fig8}%
(e)], pentagon [Figs.~\ref{fig8}(b) and \ref{fig8}(f)], hexagon [Figs.~%
\ref{fig8}(c) and \ref{fig8}(g)] and octagon [Figs.~\ref{fig8}(d) and \ref{fig8}(h)] boundaries. Note that, on the Penrose tiling QL, we actually realize an extrinsic higher-order topological insulator~\cite{Geier2018PRB} as the combined spatial symmetries $C_{4,8}m_z$ that provide topological protection in the AB tiling case are absent here. The location of the CSs can also be explained by the Jackiw-Rebbi mechanism.

\section{V. Quasicrystalline QI on the Penrose tiling QLs}
\begin{figure*}[pthb]
\centering
\includegraphics[width=12cm]{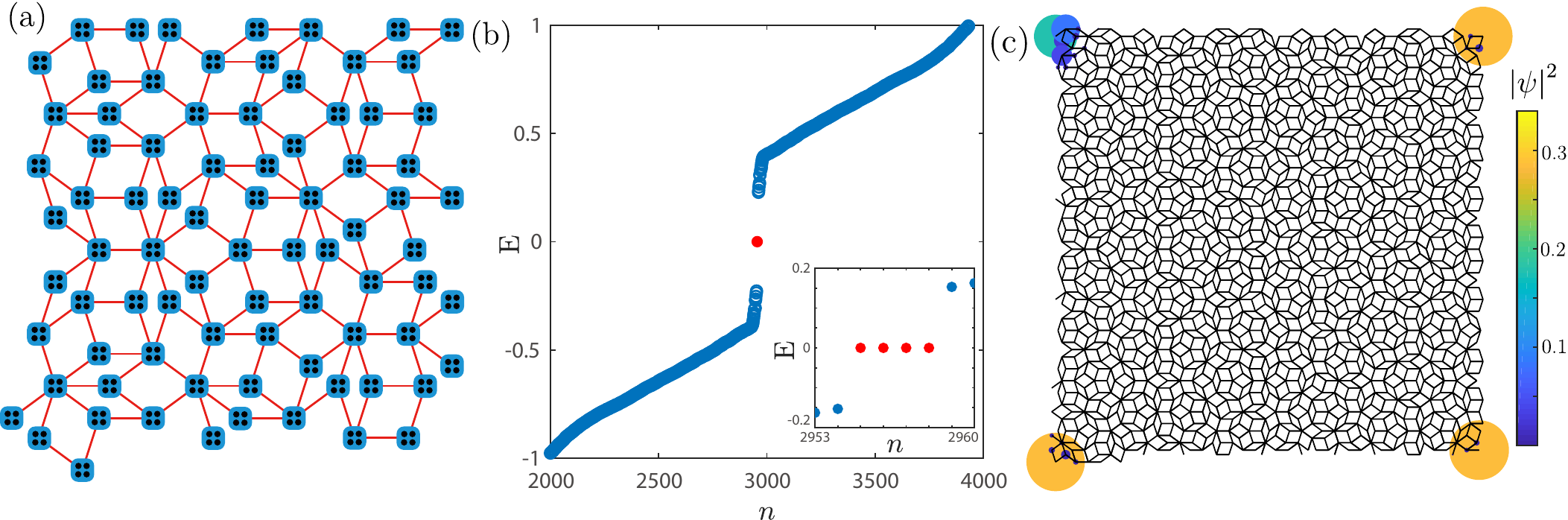}
\caption{(a) Schematic drawing of the Penrose tiling QL. Shaded blue squares
mark the cells arranged in the QL. (b) The spectra of the QI model $H_\text{%
QI}$ on the Penrose tiling QL with a square boundary. The insets denote the
zoomed-in section of ZEMs marked as the red dots. (c) The wavefunction
probability of the ZEMs.}
\label{fig9}
\end{figure*}

In the main text, we have presented the numerical results of the QI on the
AB tiling QL. Actually, the QI can be realized on the Penrose tiling QL as
well. The designed Penrose tiling QL is shown in Fig.~\ref{fig9}(a). We
use four sites to form a cell, and the cells are arranged in a QL according
to the Penrose tiling. After diagonalizing the Hamiltonian $H_\text{QI}$%
~[Eq.~(3) in the main text.], we find four in-gap ZEMs~[see Fig.~\ref{fig8}%
(b)] localized at the four corners~[see Fig.~\ref{fig9}(c)].

\section{VI. Quasicrystalline QI with the simplified inter-cell hopping}
\begin{figure}[h]
\centering
\includegraphics[width=8cm]{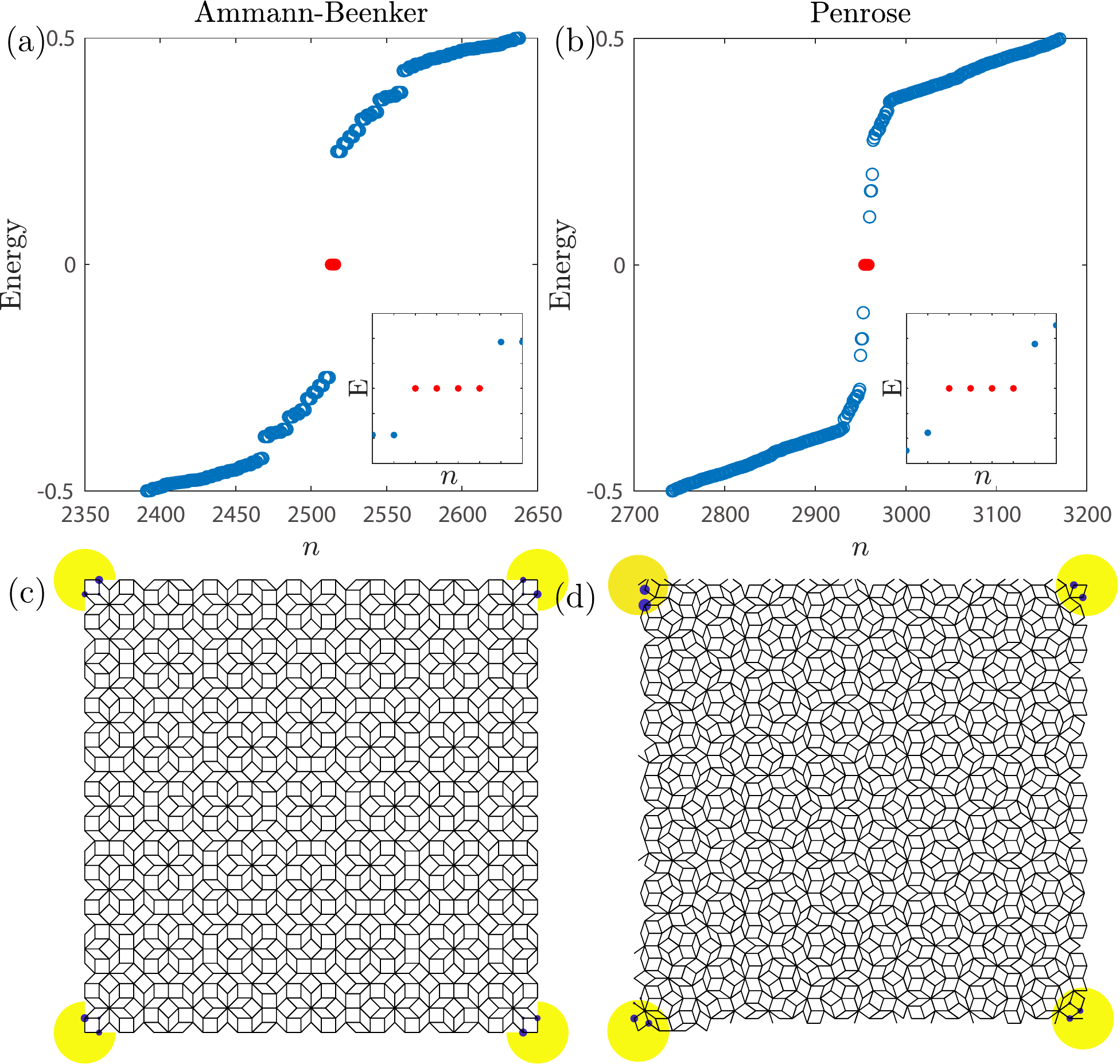}
\caption{The energy spectra of the QI model $H_\text{QI}$ with the
simplified $T(\protect\phi_{jk})$ on the AB tiling~(a) and the Penrose
tiling~(b) QLs with a square boundary. The insets denote the zoomed-in
section of ZEMs marked as the red dots. (c) and (d) The wavefunction
probability of ZEMs on the AB and Penrose tiling QLs.}
\label{fig10}
\end{figure}

To design a simple EC to engineer the QI in quasicrystals, we make a
simplification about the inter-cell hopping $T\left( \phi_{jk}\right)$ in
the Hamiltonian $H_\text{QI}$. We find that, both the AB tiling QL and the
Penrose tiling QL we design can still host the QI state for the simplified
inter-cell hopping. To demonstrate this, we present the numerically
calculated spectra and the wavefunction probability of ZEMs in Fig.~\ref%
{fig10}. Under open boundary condition, four CSs appear at the four corners
of the QLs.

\section{VII. More information about the designed EC to simulate the quasicrystalline
QI}

In our designed EC, nodes are connected by capacitors or inductors according
to the sign of hoppings. The circuit cells are placed at the sites of the
QL. Each circuit cell consists of four nodes~(sites). The nodes within a
circuit cell are connected by a pair of capacitor $C_0$ and inductor $L_0$.
The nodes between different circuit cells are coupled by pairs of $C_{1,2,3}$
and $L_{1,2,3}$. The capacitances satisfy the relation $C_{m}=e^{1-r_{m}/%
\xi}C_{0}\lambda /\gamma$ and the inductivities are $L_{m}=L_{0}\gamma
/\left( e^{1-r_{m}/\xi}\lambda \right)$ with $m=1,2,3$. In this way,
connections in the EC show the same spatial dependency as the hopping
integrals in the Hamiltonian $H_\text{QI}$ on the QL.

As stated in the main text, the EC Hamiltonian $H$ and the QL Hamiltonian $H_%
\text{QL}$ are related through the relation $H(\omega)=H_\text{QL}%
(\omega)+D(\omega)$. Here $D(\omega)$ is a diagonal matrix contains all the
diagonal elements in $H(\omega)$. In other words, these two Hamiltonian are
equivalent except for the diagonal matrix $D$. To observe the CS of $H_\text{%
QL}$, we need to set $D(\omega)=0$. This can be done by a suitable choice of
the grounding at the resonance frequency $\omega_0$. In this situation, the
impedance $Z(\omega_0) \propto 1/j_l(\omega_0)$ diverges when there exists a
CS, which gives rise to $j_l(\omega_0)=0$. In the designed EC, the pairs of $%
C_{0,1,2,3}$ and $L_{0,1,2,3}$ are chosen to have the same resonance
frequency \textit{i}.\textit{e}., $\omega_0=1/\sqrt{C_0L_0}=1/\sqrt{C_1L_1}%
=1/\sqrt{C_2L_2}=1/\sqrt{C_3L_3}$. Note that, in the quasicrystalline EC, unlike
in the periodic ECs, the grounding for nodes in each circuit cell are
different. In our numerical calculation, for a given $\omega_0$, we choose
the grounded $C_{pa,g}$ and $L_{pa,g}$ at node $pa$ according to $C_{pa,qa}=%
\frac{1}{\omega^2_0}W_{pa,qa}$, where $C_{pa,pa}=-C_{pa,g}-\sum_{q^{\prime
}b^{\prime }}C_{pa,q^{\prime }b^{\prime }}$ and $W_{pa,pa}=-L^{-1}_{pa,g}-%
\sum_{q^{\prime }b^{\prime }}L^{-1}_{pa,q^{\prime }b^{\prime }}$~\cite%
{Imhof2018NatPhys}.